\documentclass[aps,prb,groupaddress,twocolumn,showpacs,superscriptaddress]{revtex4-1} 

\usepackage[colorlinks=true, linkcolor=blue, citecolor=blue, urlcolor=blue]{hyperref}
\usepackage{amsmath,amssymb,mathrsfs}
\usepackage{latexsym}
\usepackage{graphicx} 
\usepackage{epstopdf}
\usepackage{graphicx,epstopdf,color}
\usepackage{amsfonts}
\usepackage{bm}
\usepackage{bbm}
\usepackage{multirow}
\usepackage{rotating}
\usepackage{overpic}

\graphicspath{./}
\definecolor{forestgreen}{rgb}{0.13, 0.60, 0.13}

\def\ie{\emph{i.e.},\ }
\def\eg{\emph{e.g.}\ }

\newcommand{\ket}[1]{\left|#1\right\rangle}

\newcommand{\up}{\uparrow}
\newcommand{\dw}{\downarrow}
\newcommand{\pd}{{\phantom{\dagger}}}
\newcommand{\ps}{{\phantom{*}}}

\begin{document}

%\title{Doping a topological insulator:\\
%a promising strategy to find topological superconductors?}
\title{Topological superconductivity on the honeycomb lattice:\\
Effect of normal state topology }

\author{Sebastian Wolf}
\affiliation{School of Physics, University of Melbourne, Parkville, VIC 3010, Australia}
\author{Tyler Gardener}
\affiliation{School of Physics, University of Melbourne, Parkville, VIC 3010, Australia}
\author{Karyn Le Hur}
\affiliation{CPHT, CNRS, Institut Polytechnique de Paris, Route de Saclay, 91128 Palaiseau, France}
\author{Stephan Rachel}
\affiliation{School of Physics, University of Melbourne, Parkville, VIC 3010, Australia}
 \pagestyle{plain}

\begin{abstract}
The search for topological superconductors is one of the most pressing and challenging questions in condensed matter and material research. Despite some early suggestions that {\it doping a topological insulator} might be a successful recipe to find topological superconductors, until today there is no general understanding of the relationship of the topology of the superconductor and the topology of its underlying normal state system. One of the major obstacles is the strong effect of the Fermi surface and its subsequent pairing tendencies within the Hubbard model framework, usually preventing a detailed comparison between different topological superconducting systems. Here we present an analysis of doped insulators--topological and trivial--where the dominant Fermi surface effects have been equalized. Our approach allows us to study and compare superconducting instabilities of different normal state systems and present rigorous results about the influence of the normal state system's topology.
%reveal the potential of doping a topological insulator.
\end{abstract}

%\date{\today}

\maketitle

\section{Introduction} 
\label{sec:Intro}
%
%{\it Introduction.---} 
Topological superconductors (TSCs)\,\cite{sato_topological_2017} are among the most desirable materials since their Majorana quasi particles exhibit Non-Abelian statistics\,\cite{ivanov01prl268} and are considered the prime candidates for topological quantum computing\,\cite{nayak-08rmp1083}. After searching for TSCs for a few decades, we still have not established a general method or a ``recipe'' of how to fabricate or synthesize them, despite some promising strategies\,\cite{ramires-18prb024501,sato09prb214526,sato10prb220504,fu-10prl097001,geier_symmetry-based_2020,ono_refined_2020}.
% sato09prb214526,fu-10prl097001,sato10prb220504, symmetry indicator papers by japanese guys}. 
An exception constitute hybrid and engineered systems, where due to proximitization of $s$ wave superconductivity the system's ground state can effectively become topologically non-trivial\,\cite{fu-08prl096407,lutchyn-10prl077001,oreg-10prl177002,mourik_signatures_2012,nadj-perge-14s602,palacio-morales_atomic-scale_2019}. %,crawford-21arXiv}.

Here we focus on the key question what it takes to make unconventional superconductors -- driven by electron correlations -- topologically non-trivial. These {\it intrinsic} TSCs are often also referred to as spin-triplet or odd-parity superconductors. More generally, also gapped spin-singlet superconductors can be considered topologically non-trivial when they are chiral\,\cite{kallin_chiral_2016}, as \eg realized through $d+id$ pairing on hexagonal lattices\,\cite{nandkishore-12np158,kiesel-12prb020507,wu-13prb094521,black-schaffer-14prb054521,black-schaffer-14jpcm423201}; chiral singlet pairing leads, however, to an even-valued Chern number and an even number of Majorana zero modes at defects, allowing them to annihilate into trivial states\,\cite{footnote-sato}.
Examples of candidate materials for odd-parity pairing which have attracted much attention lately are UPt$_3$\,\cite{tou-96prl1374}, CePt$_{3}$Si\,\cite{bauer-04prl027003,yanase-08jpsp124711}, Cu$_x$Bi$_2$Se$_3$\,\cite{hor-10prl057001,fu-10prl097001,sasaki-11prl217001}, 
Sn$_{1-x}$In$_x$Te\,\cite{erickson-09prb024520,novak-13prb140502,sasaki-12prl217004},
FeSe$_{0.45}$Te$_{0.55}$\,\cite{zhang-18s182,rameau-19prb205117,wang-20s104} and UTe$_2$\,\cite{ran-19s684}.
Amongst them, Cu$_x$Bi$_2$Se$_3$ and Sn$_{1-x}$In$_x$ are of particular interest, since both their normal state systems represent doped topological materials: Bi$_2$Se$_3$ is a three-dimensional topological insulator (TI)\,\cite{zhang-09np438} and SnTe a topological crystalline insulator\,\cite{tanaka-12np800}. That raises the question whether it might be advantageous to search for TSCs in doped topological (insulating) materials. 
This central question has occurred since the discovery of TIs itself, since the Bloch matrix of TIs and the Bogoliubov--de Gennes matrix of TSCs can be form-equivalent. The simplest case to illustrate this, are the two-orbital Chern insulator and the chiral $p$ wave superconductor on the square lattice. Both two-dimensional matrices contain $\pm2t [\cos{(ap_x)}+\cos{(ap_y)}]$ on the diagonal: $+$ ($-$) corresponds to the electron (hole) degree of freedom of the superconductor or to the single particle dispersions of the two different orbitals, respectively. The off-diagonal element is of the form $\sin{(ap_x)} + i \sin{(ap_y)}$ and realizes chiral $p_x + i p_y$ pairing, or the complex hybridization between the two orbitals responsible for the band inversion of the Chern insulator. Does this analogy imply that doped topological insulators possess an advantage in realizing topological superconductivity? An affirmative answer is not possible for two reasons:
%The answer is clearly negative for two reasons: 
(i) (Topological) superconductivity %, and thus also topological superconductivity, 
emerges for energetic reasons, while topological states of matter such as TIs for symmetry reasons. (ii) TIs in their simplest form are non-interacting systems; in contrast, TSCs in the Bogoliubov--de Gennes framework are only treated on the mean-field level. Their full description requires, however, to include the electron--electron interactions of the many-body Hubbard model. 
%A consequence of the latter point is that a topological bandstructure would only affect the diagonal blocks of the Bogoliubov--de Gennes matrix; the off-diagonal blocks which contain the pairing need to be derived from the Hubbard interactions in an unbiased way. 
As a consequence, it is {\it a priori} unknown what the pairing and the topology of the resulting superconducting groundstate might be. 
%*although some indicators were formulated recently\,\cite{ramires-18prb024501}).
This situation calls for a systematic investigation of Hubbard models where the kinetic term can or cannot be topologically non-trivial. It is well-known, however, that already small changes of the Fermi surface (FS) can drastically influence which superconducting instability prevails; information about the topology of the normal state system is only stored in the eigenvectors of the Bloch matrix, due to the dominance of FS effects it is often difficult or even impossible to undertake such a systematic investigation.

%\newpage

%\begin{itemize}
% \item topological insulators / superconductors: very interesting (Majoranas, quantum computing etc.)
% \item SC in doped topological insulators (references)
% \item SC in hexagonal lattice\,\cite{nandkishore_chiral_2012,nandkishore_superconductivity_2014}
% \item Method: WCRG: highly sensitive to the shape of the FS $\Rightarrow$ dope lattice models with different topology but same FS and see what the SC instability does
%\end{itemize}
%
%Use this:\\
%``Our key result, which is also emphasized in the conclusion, is the observation that already the inclusion of a second orbital into the minimal model yields a broad range of possible instability scales due to the complicated Fermiology and orbital makeup of the quasiparitcle excitations.''

In this Letter, we study the superconducting states of topologically different models on the honeycomb lattice. We carefully choose the system parameters such that their FSs are {\it identical}, thus equalizing the effect of the FS. The remaining main difference between these models, entering the calculation of the superconducting ground state, are their eigenvectors which store the information about the normal state system's topology. We apply the weak coupling renormalization group (WCRG) approach\,\cite{raghu_superconductivity_2010,raghu_effects_2012,wolf_unconventional_2018,cho-13prb064505,platt-16prb214515,cho-15prb134514,scaffidi-14prb220510,roising-18prb224515} which limits us to infinitesimal repulsive interactions; however, the method is asymptotically exact.

%%%%%%%%%%%%%%%%%%%%%%%%%%%%%%%%%%%%%%%%%%%%%%%%%%%%%%%%%%%%%%%%%
%
%                M O D E L    A N D    M E T H O D
%
%%%%%%%%%%%%%%%%%%%%%%%%%%%%%%%%%%%%%%%%%%%%%%%%%%%%%%%%%%%%%%%%%

\section{Models and methodology} 
\label{sec:model}
%
%{\it Models and methodology.---}
We study superconducting instabilities of four different bandstructures on the honeycomb lattice with additional Hubbard onsite interaction, realizing the simplest model to give an answer to the question of this Letter. %Since our interest is in doped TIs, %we consider two topological and two trivial insulators on the honeycomb lattice. The former are the spinful Haldane (Hd) model\,\cite{haldane88prl2015}, a time-reversal breaking Chern insulator, and the Kane-Mele (KM) model\,\cite{kane-05prl146802,kane-05prl226801}, a time-reversal invariant $\mathbb{Z}_2$ TI. 
Specifically, we consider as a topologically non-trivial system on the honeycomb lattice the Kane-Mele (KM) model\,\cite{kane-05prl146802,kane-05prl226801}, a time-reversal invariant $\mathbb{Z}_2$ TI. 
%The latter are the
In addition, we study the inversion symmetry breaking Semenoff (Se) insulator\,\cite{semenoff84prl2449}, realized through a spin-independent staggered sublattice potential, and the antiferromagnetic Zeeman (AF) insulators\,\cite{black-schaffer-15prb140503}, realized through an antiferromagnetic out-of plane Zeeman field as examples of topologically trivial insulators. These three models are complemented by the semi-metallic tight-binding model with nearest- and second-nearest neighbor hopping, referred to as graphene (Gr) in the following. The Hamiltonian is given by $\hat{H}=\hat{H}_{0}+\hat{H}_{\text{int}}$ with
\begin{equation}\label{eq:hamiltonian}
\begin{split}
% \hat{H}&=\hat{H}_{0}+\hat{H}_{1}+\hat{H}_{\text{int}}\ ,\nonumber\\[5pt]
 \hat{H}_{0}&=\sum_{i,j}\sum_{s,s'}c_{is}^{\dagger}\hat{h}_{ij,ss'}^{\pd}c_{js'}^{\pd}\ ,\\
 \hat{H}_{\text{int}}&=U_{0} \sum_{i}\sum_{s\neq s'}c_{is}^{\dagger}c_{is'}^{\dagger}c_{is'}^{\pd}c_{is}^{\pd}\ ,
 \end{split}
 \end{equation}
where $\hat{h}$ denotes the $4\times 4$ hopping matrix, $c_{is}^{\dagger}$ is the creation operator of an electron with spin $s$ at site $i$, and $U_{0}$ is the onsite interaction strength. The Bloch matrix is defined for each of the four models in Sec.\,II of the Supplement\,\cite{supp}; Fig.\,\ref{fig:models} contains a visual representation of all hopping and onsite terms. We note that the energy bands have the general form 
\begin{equation}
 E(\vec{k},n)=\,\Lambda_{0}(\vec{k})+n\sqrt{T_{1}^{2}(\vec{k})+T_{2}^{2}(\vec{k})+\Lambda_{1}^{2}(\vec{k})} \nonumber
 \end{equation}
 with the band index $n=\pm1$.
 % and $\vec k$ dependent functions $\Lambda_{0}$, $T_1$, $T_2$, and $\Lambda_1$;  
 % defined in \sr{Sec.\,III} in the Supplement. 
For simplicity the details of the $\vec k$ dependent functions functions $\Lambda_{0}$, $T_1$, $T_2$, and $\Lambda_1$ are shown in Sec.\,III of the Supplement\,\cite{supp}.
We stress that $\Lambda_0$ is non-zero only for the Gr model; for all other models the energy bands exhibit a particle-hole symmetry and it is sufficient to focus w.l.o.g.\ on the electron-doped case (\ie filling $n>1$).

%%%%%%%%%%%%%%%%%%%%%%%%%%%%%%%%%%%%%%%%%%%%%%%%%%%%%%%%%%%%%%
%           F I G. 1
%%%%%%%%%%%%%%%%%%%%%%%%%%%%%%%%%%%%%%%%%%%%%%%%%%%%%%%%%%%%%%
\begin{figure}[t!]
 \includegraphics[width=0.70\columnwidth]{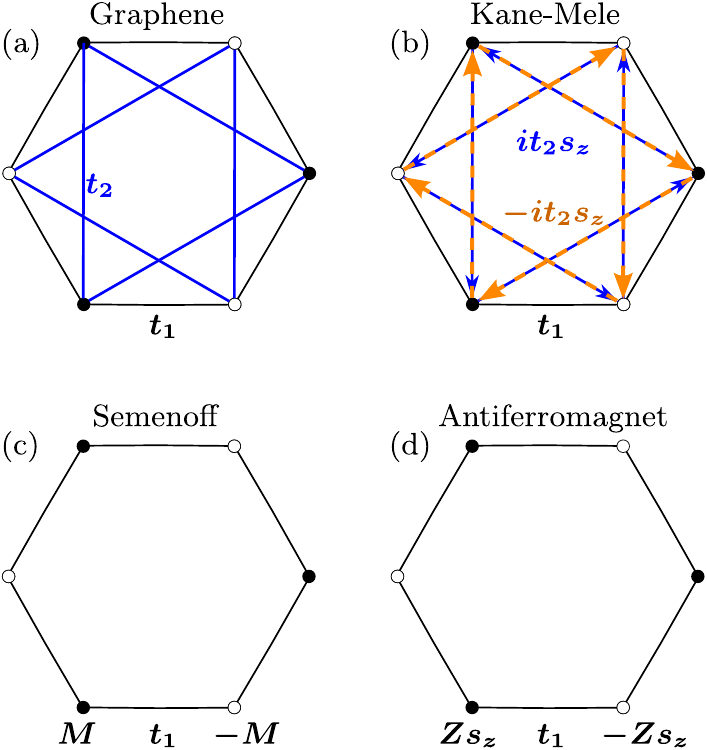}
 \caption{\label{fig:models}
 Hoppings and onsite potentials of the four honeycomb lattice systems studied in this paper. Atoms on sublattice $A$ ($B$) are indicated by full (empty) circles.
 a) Graphene model with nearest and next-nearest neighbor hoppings. 
 b) Kane-Mele model, where the next-nearest neighbor hopping is imaginary, with opposite sign for clockwise (orange) and counter-clockwise (blue) hopping; in addition, there is a sign change for different spins.
 c) Semenoff model with staggered onsite potential $M$. 
d) Antiferromagnetic Zeeman model, which is similar to Semenoff model with the difference that the onsite potential $Z$ changes sign for different spins.
 %d) Haldane model, where the next-nearest neighbor hopping is imaginary, with opposite sign for clockwise (orange) and counter-clockwise (blue) hopping. 
 } 
\end{figure}
%%%%%%%%%%%%%%%%%%%%%%%%%%%%%%%%%%%%%%%%%%%%%%%%%%%%%%%%%%%%%%
%%%%%%%%%%%%%%%%%%%%%%%%%%%%%%%%%%%%%%%%%%%%%%%%%%%%%%%%%%%%%%
%           T A B. 2
%%%%%%%%%%%%%%%%%%%%%%%%%%%%%%%%%%%%%%%%%%%%%%%%%%%%%%%%%%%%%%
\begin{table}[b!]
 \caption{\label{tab:symmetries}Symmetries of the Bloch matrix of the four different models. TR denotes time reversal symmetry, I denotes inversion symmetry, $C_{n}$ are rotations and $M_{i}$ reflections.}
 \centering
 \begin{tabular}{c|cc|ccccc|l}
  \hline
   & & & \multicolumn{5}{c|}{lattice symmetries} & residual \\
  Model & TR & I & $C_{2}$ & $C_{3}$ & $C_{6}$ & $M_{x}$ & $M_{y}$ & group  \\
  \hline\hline
  Gr & X & X    & X & X & X & X & X & $D_{6}$ \\
  KM & X & X    & X & X & X & X & X & $D_{6}$ \\
  Se & X &      &   & X &   &   & X & $D_{3}$ \\
  AF &   &      &   & X &   & X &   & $D_{3}$ \\
 % Hd &   & X    & X & X & X &   &   & $C_{6h}$ \\
    \hline
 \end{tabular}
\end{table}
%%%%%%%%%%%%%%%%%%%%%%%%%%%%%%%%%%%%%%%%%%%%%%%%%%%%%%%%%%%%%%

In the following, we emphasize some basic facts about superconductivity and the specific aspects on methodology which are key for this work. 

While conventional, phonon-mediated superconductors form Cooper pairs with zero angular momentum ($s$ wave), unconventional superconductors usually involve higher-angular momentum pairing channels (such as $p$, $d$, $f$ wave). It is well-established that the crystal space group of the normal state dictates the possible irreducible representations (irreps), and the superconducting order parameter must transform according to one of these irreps\,\cite{annett90ap83}. The symmetry analysis of the four considered models is summarized in Tab.\,\ref{tab:symmetries}. Gr and KM correspond to symmetry group $D_6$ %, Hd to $C_{6h}$ 
and Se and AF to $D_3$. We note that the differences are small: $D_3$ only consists of the three irreps $A$, $B$ and $E$, while they are split for $D_6$ %and $C_{6h}$ 
into $A_1$ and $A_2$, $B_1$ and $B_2$ as well as $E_1$ and $E_2$. Most notably are the two-dimensional irreps $E_1$ containing the $p$ wave and $E_2$ the $d$ wave order parameters, which can form chiral superpositions such as $p_x + i p_y$ and $d_{xy} + i d_{x^2-y^2}$. For $D_3$ these belong to the same irrep and could possibly be combined (which we have not observed here).
We further note that the AF and KM models further break the SU(2) spin rotation symmetry down to only U(1) symmetry (\ie spin is still a good quantum number) which can lead to a splitting of the three degenerate triplet states into $S^z=0$ and $\pm 1$ components. 
%\sr{Last but not least we emphasize that the Hd model can enable Cooper pair wavefunctions where in addition to spin and angular momentum also the sublattice degree of freedom becomes active; as a consequence exotic condensates such as triplet $d$ wave and singlet $p$ wave are possible - which we indeed observe for some parameter ranges, see below.}

\begin{figure}[t!]
 \centering\includegraphics[width=0.8\columnwidth]{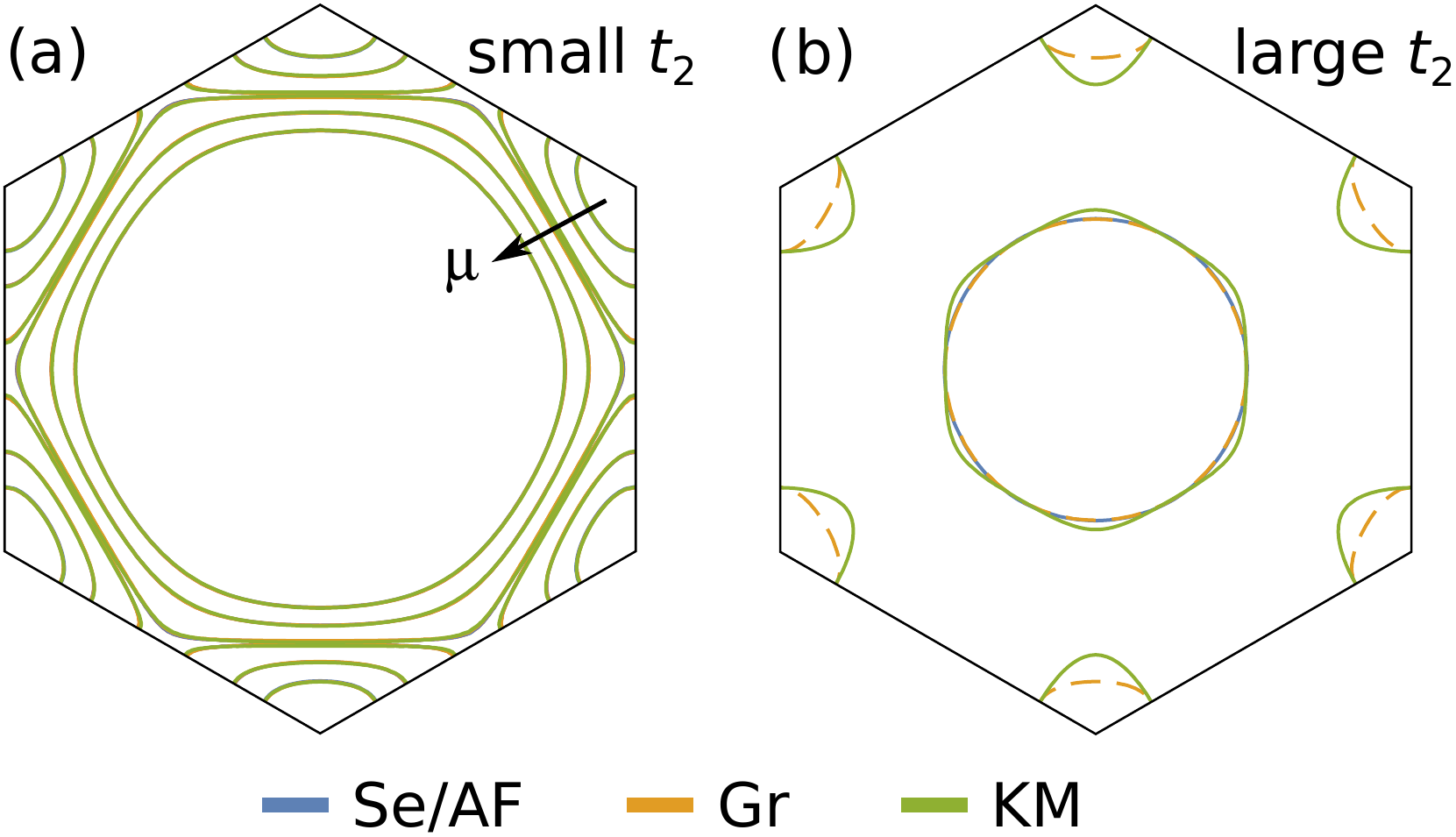}
 \caption{\label{fig:FSs}(a) Fermi surfaces of the different models for small values of $t_{2}$ and different values of $\mu$. We note that $t_2$ refers here either to the second neighbor hopping or to the onsite terms, respectively. For clarity, we only show a subset of all used values of $\mu$: The six FSs closest to the corners of the BZ correspond to $\mu=0.6$, and the almost circular FS in the center of the BZ to $\mu=1.4$. %The arrow indicates the direction in which $\mu$ increases. 
 The specific values of $t_2$ are listed in the Supplement for all values of $\mu$\,\cite{supp}.
  %For each value of $\mu$, the values of $t_{2}$ for each model can be adjusted to obtain (almost) identical Fermi surfaces. 
 (b) Fermi surfaces of the different models for large values of $t_{2}$. We only show an example of Fermi surfaces which consist of multiple pockets, one in the center and additional ones in the corner of the BZ (KM and Gr models). Se and AF models only possess a single circular FS.} 
\end{figure}

Superconductivity is a weak-coupling instability of the Fermi sea, thus small changes of single-particle parameters, which lead to changes of the FS, can have drastic effects on the symmetry of the superconducting ground state. In particular, small changes of the FS can trigger a phase transition between superconducting ground states which transform according to different irreps.  We note that the applied WCRG is particularly sensitive to changes in the FS due to the vanishingly small interaction scale; we stress, however, that the same is true for other methods designed to work at intermediate coupling strength such as random phase approximation\,\cite{altmeyer_role_2016} or functional renormalization group\,\cite{metzner_functional_2012}. That is why it does not only make little sense to compare topological bandstructures on different lattice geometries, but neither is it very meaningful to compare them when their FSs are different. The honeycomb lattice with its low-energy Dirac bands offers the perfect setting to overcome both obstacles. Small-gap insulating phases are usually well characterized by a gapped Dirac Hamiltonian as an effective low-energy theory at both valleys of the honeycomb lattice. To lowest order, the gap term $\propto G$ is $\vec k$ independent, $H_{\rm Dirac} \sim k_x \sigma^x + k_y \sigma^y + G \sigma^z$. Thus it is clear that for small gap opening $\sim G$ and not too large doping, different insulating honeycomb lattice bandstructures can produce identical FSs. We show in Fig.\,\ref{fig:FSs}\,(a) that even for large electron doping FSs remain identical if parameters are carefully chosen. Moreover, we are able to adjust the second neighbor hopping $t_2$ and the filling $n$ of the Gr model to match these FSs as well [see Fig.\,\ref{fig:FSs}\,(a)].

%
%%%%%%%%%%%%%%%%%%%%%%%%%%%%%%%%%%%%%%%%%%%%%%%%%%%%%%%%%%%%%%
%           F I G. 2
%%%%%%%%%%%%%%%%%%%%%%%%%%%%%%%%%%%%%%%%%%%%%%%%%%%%%%%%%%%%%%
%%%%%%%%%%%%%%%%%%%%%%%%%%%%%%%%%%%%%%%%%%%%%%%%%%%%%%%%%%%%%%
%

As a method we employ the WCRG which is a realization of Kohn-Luttinger superconductivity\,\cite{kohn_new_1965}. Details about the WCRG are delegated to the Supplement\,\cite{supp} (see, also, references\,\cite{wolf_spin-orbit_2020,vafek_spin-orbit_2011} therein). Here we only discuss the structure of the analytical expression of the two-particle vertex $\Gamma$ which is the central object we compute. In second order of the local electron-electron intraction $U_0$ we have  $\Gamma^{(2)}=\,\frac{1}{2}\Gamma_{\text{BCS}}+\Gamma_{\text{ZS}}+\Gamma_{\text{ZS'}}$.
 The first contribution is referred to as the BCS diagram and the other two contributions as zero sound (ZS) diagrams\,\cite{shankar_renormalization-group_1994}. The BCS diagram diverges logarithmically, which indicates a phase transition, \ie that the normal state becomes unstable towards the superconducting state (superconductivity is the only instability in the weak coupling regime\,\cite{kohn_new_1965}). The divergency is treated by renormalization, upon which we obtain the superconducting order parameter from the eigenvalue problem of the remaining two contributions $\Gamma_{\text{ZS}}$ and $\Gamma_{\text{ZS'}}$. They take the form
 \begin{equation}\label{2v}\begin{split}
 \Gamma_{\rm ZS}(2\bar{2}'\bar{1}'1)=\sum_{\vec{k}_{3}}&\sum_{n_{3}n_{4}}\sum_{\tilde{s}_{3}\tilde{s}_{4}}M(\bar{2}'431)M(234\bar{1}')\\
 &\times\frac{f(E(3))-f(E(4))}{E(3)-E(4)}\ ,
\end{split}\end{equation}
with the Fermi distribution $f(E)$. We use the short notation  $1\equiv \vec{k}_{1},n_{1},\tilde{s}_{1}$, $\bar{1}\equiv-\vec{k}_{1},n_{1},\tilde{s}_{1}$, $1'\equiv \vec{k}_{1},n_{1},\tilde{s}_{1}'$. The expression \eqref{2v} has essentially three ingredients: (i) The FS enters via incoming and outgoing momenta, $\vec k_1$ and $\vec k_2$, respectively, and has a strong influence on the value of \eqref{2v}, as mentioned before. (ii) At zero temperature, the Lindhard function (the expression in the second line) contains the energy bands, albeit suppressed with $1/E$. In particular, here also energy states enter which are not at the FS. We have compared the energy bands of all four models, and the differences between them are very small or negligible; obviously they are biggest where Gr is gapless while all other models are gapped. Nonetheless, we expect the Lindhard expression to be essentially the same for all models, when their FSs are identical (see Sec.\,VI in \,\cite{supp}). (iii) The third ingredient is the product of $M$ factors which contain the eigenvectors $\vec v$ of the Bloch matrix, containing the information about topology: $M(4321)=
  \sum_{l,s\neq s'}v_{l,s}^{*}(4)v_{l,s}^{\ps}(1)v_{l,s'}^{*}(3)v_{l,s'}^{\ps}(2) -\sum_{l,s\neq s'}v_{l,s}^{*}(4)v_{l,s}^{\ps}(2)v_{l,s'}^{*}(3)v_{l,s'}^{\ps}(1)$.   
  The main idea of this paper is to prepare the four different models such that (i) and (ii) are essentially the same, and the remaining difference is (iii).

%%%%%%%%%%%%%%%%%%%%%%%%%%%%%%%%%%%%%%%%%%%%%%%%%%%%%%%%%%%%%%%%%
%
%                                 R E S U L T S
%
%%%%%%%%%%%%%%%%%%%%%%%%%%%%%%%%%%%%%%%%%%%%%%%%%%%%%%%%%%%%%%%%%
\section{Results}
\label{sec:results}
%
%{\it Results.---}
We choose $t_{2}=0.05t_1$ for the Kane-Mele model, leading to a relatively small gap of $3\sqrt{3}t_2\approx 0.26t_1$ (in the following we refer to this as the ``small gap'' case); from an experimental perspective, this represents already a reasonable gap size. We adjust the Se and the AF models to have a comparable gap size and identical FSs; we also tune the Gr model to match the FS, although Gr of course remains gapless. Note that we need to slightly adjust these parameters whenever we change the filling fraction in order to retain identical FSs (see Tab.\,S4 in the Supplement\,\cite{supp}).

\begin{figure}[t]
 \includegraphics[width=0.8\columnwidth]{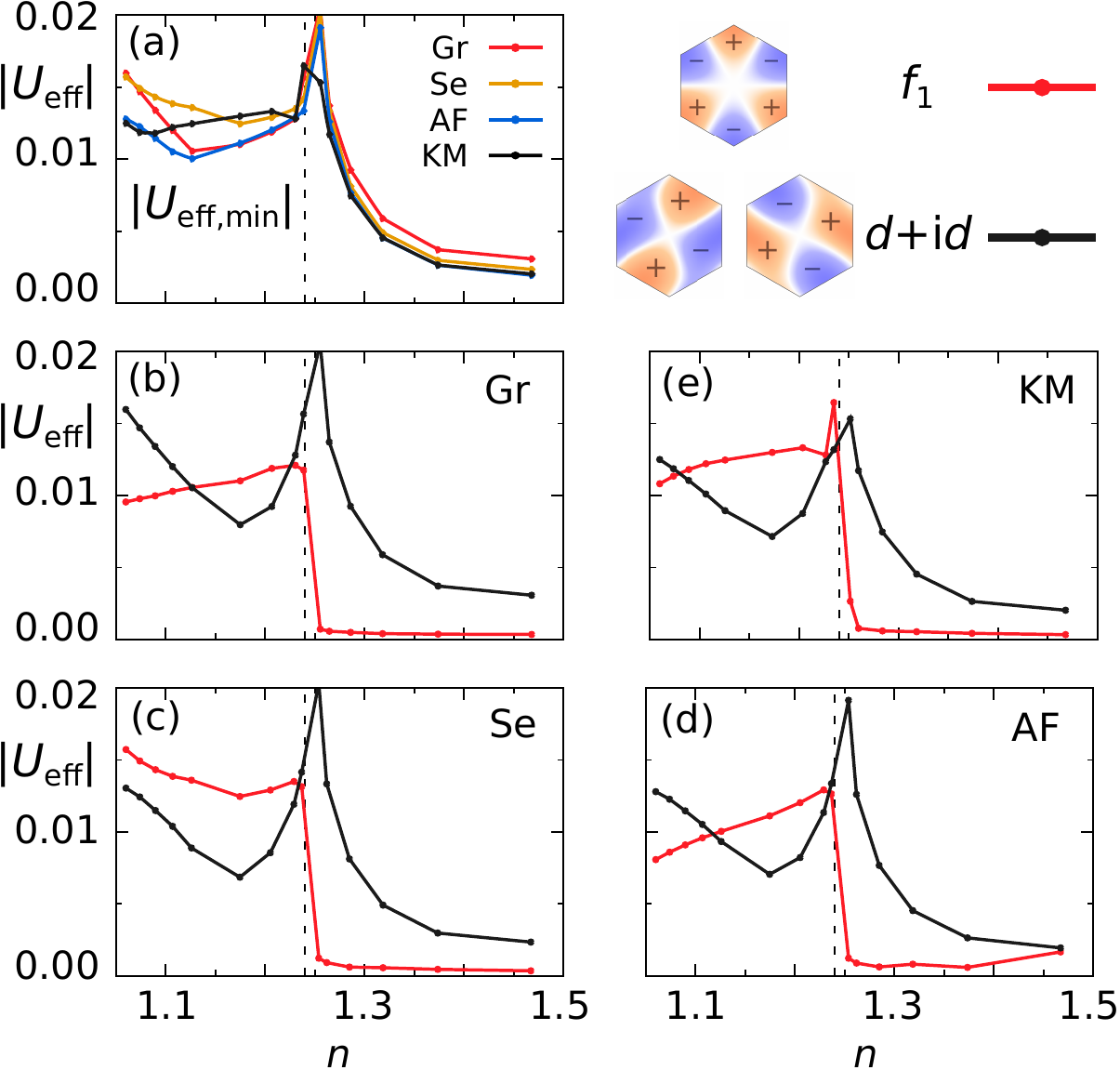}  %results_small_gap_v2.pdf} 
 \caption{\label{fig:results_small_gap}Results for the systems corresponding to the ``small gap'' case, equivalent to $t_{2}=0.05 t_1$ in the KM model.
 Filling and parameters of the other models are tuned to match the FSs of the KM model. Dashed lines indicate van Hove filling.} 
\end{figure}

By virtue of WCRG we obtain the leading superconducting instabilities for the small gap case, shown in Fig.\,\ref{fig:results_small_gap}. Surprisingly, we find for all four models chiral $d+id$ wave pairing ($E_2$ irrep) to be dominating, when the filling is higher than van Hove filling (at $\mu/t_1\approx 1$ corresponding to $n\approx 1.24$). Below van Hove filling, we find for all four models $f$ wave pairing ($B_1$ irrep) as the leading instability, referred to as $f_1$ in the following. Only for even further reduced fillings, $d+id$ pairing wins once more (with the exception of the Semenoff model). Despite these minor quantitative differences, we find qualitatively  the same phases in all models in similar ranges of $n$.
%\footnote{Since all models exhibit a particle-hole symmetric single-particle spectrum (or almost particle-hole symmetric in the case of Gr), superconducting solutions for the hole-doped case are the same as for the electron-doped case.}. 
We note that both chiral $d$ wave and $f$ wave result here in a fully gapped excitation spectrum.
Fig.\,\ref{fig:results_small_gap}\,(a) shows the effective coupling strength for all models, which is roughly the same. 
We note that all models are particle-hole symmetric except Gr, and thus our results for the hole-doped case are the same as those shown in Fig.\,\ref{fig:results_small_gap}; here also for Gr the results do not change qualitatively for hole-doping\,\cite{wolf_unconventional_2018}.
Panels (b)--(e) show the leading instabilities of each model individually.
That is the key result of this Letter: we do not observe any relevant influence of the normal-state system's topology onto the superconducting ground state for the considered models.

%For all models, we see either and $f$-wave (B$_1$) or a $d+id$-wave (E$_2$) solution across the whole range of chemical potentials studied. In the hole doped regime, \ie for $\mu\lesssim1$, there are small quantitative differences for the effective interaction between the models. However, qualitatively, we find the same phases in all models in similar ranges of $\mu$.
%
%%%%%%%%%%%%%%%%%%%%%%%%%%%%%%%%%%%%%%%%%%%%%%%%%%%%%%%%%%%%%%
%           F I G. 3
%%%%%%%%%%%%%%%%%%%%%%%%%%%%%%%%%%%%%%%%%%%%%%%%%%%%%%%%%%%%%%
%%%%%%%%%%%%%%%%%%%%%%%%%%%%%%%%%%%%%%%%%%%%%%%%%%%%%%%%%%%%%%
%
%
%%%%%%%%%%%%%%%%%%%%%%%%%%%%%%%%%%%%%%%%%%%%%%%%%%%%%%%%%%%%%%
%           F I G. 4
%%%%%%%%%%%%%%%%%%%%%%%%%%%%%%%%%%%%%%%%%%%%%%%%%%%%%%%%%%%%%%
\begin{figure}[t!]
 \includegraphics[width=0.8\columnwidth]{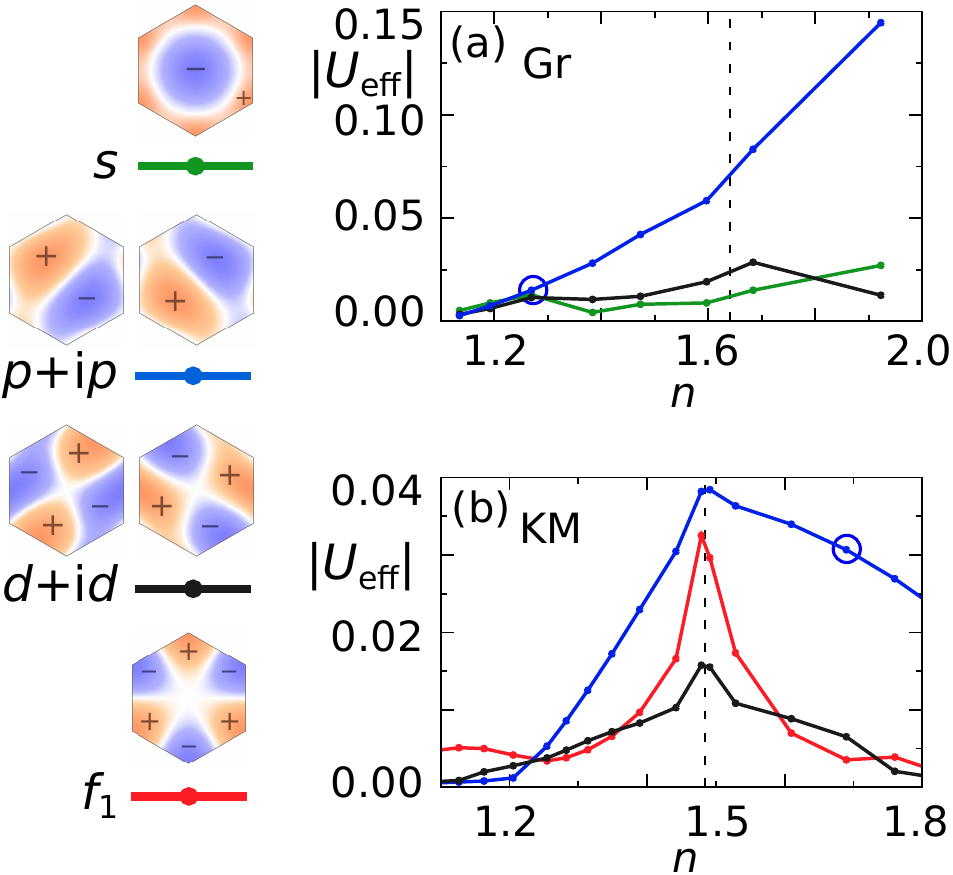} %  results_large_gap_v2.pdf}
 \caption{\label{fig:results_large_gap}Results for KM model with large gap ($t_2=0.5\,t_1$) and Gr model ($t_2=0.35\,t_1$) to feature a similar FS.
 The blue circles indicate where the Fermi surfaces in both shown models are still similar. The respective Fermi surfaces are shown in Fig.\,\ref{fig:FSs}\,(b). Dashed lines indicate van Hove filling.} 
\end{figure}
%%%%%%%%%%%%%%%%%%%%%%%%%%%%%%%%%%%%%%%%%%%%%%%%%%%%%%%%%%%%%%
%
%

Increasing the gap size would lead to a much stronger contribution of the topological (trivial) term of the KM (Se and AF) model. Due to the $\vec k$ dependence of the gap-opening term in the KM case, it is no longer possible to equalize the FSs. In fact, for a sufficiently large $t_2>0.2 t_1$ (resulting in a gap as large as $2t_1$\,\cite{rachel-10prb075106}) for the KM model, we obtain an additonal  FS pocket which is not the case for the Se and AF models, see Fig.\,\ref{fig:FSs}\,(b). At least we are able to tune the Gr model such that it mimics the FS of the KM model. In analogy to the small gap case, we refer to it as the ``large gap'' case. Fig.\,\ref{fig:results_large_gap} shows the results for Gr and KM models. Not unexpected, the superconducting phase diagrams are more diverse; they have, however, the common feature that the chiral $p$ wave instability dominates ($E_1$ irrep). 
While the $p$ wave instability for Gr is three-fold degenerate, due to the broken SU(2) symmetry the KM model only exhibits the two polarized triplet components $\ket{\up\up}$ and $\ket{\dw\dw}$ as leading instability.
The blue circle indicates the point where the FSs are still similar [Fig.\,\ref{fig:FSs}\,(b)], and indeed also the effective coupling $U_{\rm eff}$ is roughly the same. As before, the influence of the normal state's topology seems to be irrelevant. %, but we note that the signficant second-neighbor hopping seems to favor chiral $p$ wave superconductivity. 
We emphasize that both models have the common property that their FSs have pockets around the corners of the BZ in addition to the circular pocket centered around the $\Gamma$ point. Our second key result is thus that large second neighbor hopping (real or imaginary), leading to additional Fermi surface pockets around the corners of the Brillouin zone in addition to the circular pocket around the center, seems to favor chiral $p$ wave superconductivity.
That is particularly exciting as the chiral $p$ wave superconductor is known to host a Majorana zero mode at defects such as vortices or domain walls.
Considering hole-doping instead, the KM results still remain unchanged; hole-doped Gr at $t_2=0.35$ features a chiral $d$ wave state instead, due to the absence of the additional Fermi pockets at the corner of the Brillouin zone.
Last but not least we emphasize that the Gr model has a drastically increased pairing potential $U_{\rm eff}$ [see Fig.\,\ref{fig:results_large_gap}\,a)], implying that it realizes a TSC with a high critical temperature $T_c$\,\cite{wolf_unconventional_2018}. %,comment}.

Our results were derived in the weak coupling regime, and it remains unclear what might happen for larger $U_0$ and other interactions (some first steps were already taken\,\cite{lee-19prb184514}). We note, however, that a doped Kane-Mele type model was studied recently for intermediate interaction strength $U_0=2.5\,t_1$\,\cite{wu-19prb041117}, also identifying chiral $d$ wave as the leading instability due to a comparable Fermi surface with Fig.\,\ref{fig:FSs}\,a).

\section{Conclusion}
\label{sec:conclusion}
%
%{\it Conclusion.---}
We have investigated four different models on the honeycomb lattice: topological and trivial insulators as well as a variant of semi-metallic graphene. By carefully tuning the system parameters we have equalized the strong effect of the FS and of the energy bands, thus only retaining the normal state's topology as the difference between all considered models. We present rigorous results that there is {\it no} influence of the normal state's topology onto the leading superconducting instability. Instead, we find for all considered models topologically non-trivial chiral $d$ wave or odd-parity $f$ wave superconductivity.
%Thus we conclude, at least from the weak coupling perspective and on the honeycomb lattice, that {\it doping a topological insulator} into a metallic phase to find topological superconducting phases is {\it not} a promising strategy.
Clearly hexagonal lattices are a promising place to search for TSCs. Our results further suggest that longer-ranged hoppings causing additional FS pockets to appear might be an interesting avenue to find topologically non-trivial  $p$ wave superconductivity.

%In addition, 
%
%All results discussed above suggest that the topology of the superconducting phases does not depend on the topology of the underlying lattice model. Thus, from the weak coupling perspective, doping a topological insulator into a metallic phase to find topological superconducting phases is not a promising strategy. Our results indicate that what is more important to find chiral topological superconducting states is strong longer range hopping and the specific shape of the Fermi surface.
%
%
%
\begin{acknowledgments}
We acknowledge instructive discussions with R.\ Thomale and P.\ Brydon.
SR acknowledges support from the Australian Research Council through Grants No.\ FT180100211 and No.\ DP200101118. KLH acknowledges the Deutsche Forschungsgemeinschaft (DFG), German Research Foundation, under Project No.\ 277974659.
This research was undertaken using the HPC facility Spartan hosted at the University of Melbourne.
\end{acknowledgments}

\newpage

\bibliography{wcrg3}

\begin{thebibliography}{60}
\bibitem [{foo()}]{footnote-sato}%
  \BibitemOpen
  \href@noop {} {}\bibinfo {note} {In non-centrosymmetric systems it is
  possible to avoid their annihilation by applying a small Zeeman
  field\,\cite{sato-10prb134521}.}\BibitemShut {Stop}%
\end{thebibliography}

\end{document}

%\title{Supplemental material for ``Doping a topological insulator:
%a promising strategy to find topological superconductors?''}
\title{Topological superconductivity on the honeycomb lattice:\\
Effect of normal state topology }

\author{Sebastian Wolf}
\affiliation{School of Physics, University of Melbourne, Parkville, VIC 3010, Australia}
\author{Tyler Gardener}
\affiliation{School of Physics, University of Melbourne, Parkville, VIC 3010, Australia}
\author{Karyn Le Hur}
\affiliation{CPHT, CNRS, Institut Polytechnique de Paris, Route de Saclay, 91128 Palaiseau, France}
\author{Stephan Rachel}
\affiliation{School of Physics, University of Melbourne, Parkville, VIC 3010, Australia}
 \pagestyle{plain}

\maketitle

%
%%%%%%%%%%%%%%%%%%%%%%%%%%%%%%%%%%%%%%%%%%%%%%%%%%%%%%%%%%%%%%%%%
%
%                M O D E L    A N D    M E T H O D
%
%%%%%%%%%%%%%%%%%%%%%%%%%%%%%%%%%%%%%%%%%%%%%%%%%%%%%%%%%%%%%%%%%

\section{Symmetry considerations}
Before we establish the specific Bloch matrices $\hat{h}(\vec{k})$ for all the models, we start with a symmetry analysis. The hexagonal model is given by two sublattice and two spin degrees of freedom, which in total yields a 4x4 Bloch matrix. This is in general given in the following form:
\begin{align}
 \hat{h}(\vec{k})&=\sigma_{\nu}\otimes\gamma^{\nu}(\vec{k}),
\end{align}
where the $\gamma^{\nu}(\vec{k})$ are 2x2 matrices, \ie they can be written as a sum of Pauli matrices $\tau_{\mu}$ as well:
\begin{equation}
 \gamma^{\nu}(\vec{k})=\fm{\mu}{\nu}(\vec{k})\tau^{\mu}.
\end{equation}
The symmetries of the system pose several constraints on $\gamma^{\nu}(\vec{k})$, which are obtained by
\begin{equation}
 \hat{O}\hat{h}\hat{O}^{\dagger}=\hat{h},
\end{equation}
where $\hat{O}$ is the operator of the symmetry operation. Explicit representations of relevant symmetries for the hexagonal lattice are listed in Tab.\,\ref{tab:symm_op}.
%
%%%%%%%%%%%%%%%%%%%%%%%%%%%%%%%%%%%%%%%%%%%%%%%%%%%%%%
%
%       T A B.  1
%
%%%%%%%%%%%%%%%%%%%%%%%%%%%%%%%%%%%%%%%%%%%%%%%%%%%%%%
\begin{table}[b]
 \caption{\label{tab:symm_op}List of symmetry operators, $\hat{O}=\hat{O}_{s}\otimes\hat{O}_{o}\otimes\hat{O}_{r}$, for all symmetries in the $D_{6}$ group, time reversal, and inversion symmetries. $\hat{O}_{r}$ denotes the part of $\hat{O}$ which acts on real space, \ie on $\vec{r}$, $\hat{O}_{o}$ the one which acts on orbital (sublattice) space, and $\hat{O}_{s}$ the one which acts on spin space. Note that for time reversal, $\hat{T}=(\hat{T}_{s}\otimes\hat{T}_{o}\otimes\hat{T}_{r})K$, where $K$ denotes complex conjugation. $\mathbbm{1}$ is the identity (given by $\sigma_{0}$ in spin space). Note that $\hat{O}_{r}$ and $\hat{O}_{s}$ are general, while $\hat{O}_{o}$ depends on the specific arrangement of orbitals and sublattice sites in the crystal lattice.}
 %%%%%%%%%%%%%%%%%%%%%%%%%%%%%%%%%%%%%%%%%%%%%%%%%%%%%%
 \renewcommand{\arraystretch}{1.2}
 \centering\begin{tabular}{c|c|c|c}
 \hline
 Symmetry   &   $\hat{O}_{s}$ & $\hat{O}_{o}$ & $\hat{O}_{r}$   \\[0pt]
 \hline\hline
 Time reversal & $-i\sigma_{y}$  & $\tau_{0}$ &
    $\mathbbm{1}$    \\[2pt]
 Inversion      & $i\sigma_{0}$ & $\tau_{x}$ &
    $-\mathbbm{1}$  \\[4pt]
    \hline
 $C_{2}$ rotation & $\pm i\sigma_{z}$ & $\tau_{x}$ & 
    $\begin{pmatrix}
      -1 &  0 & 0 \\
       0 & -1 & 0 \\
       0 &  0 & 1 \\
     \end{pmatrix}$\\
 $C_{3}$ rotation & $\pm\Big(\frac{1}{2}\sigma_{0}+i\frac{\sqrt{3}}{2}\sigma_{z}\Big)$ & $\tau_{0}$ & 
    $\begin{pmatrix}
      -1/2 &  -\sqrt{3}/2 & 0 \\
       \sqrt{3}/2 & -1/2 & 0 \\
       0 &  0 & 1 \\
     \end{pmatrix}$\\
 $C_{6}$ rotation & $\pm\Big(\frac{\sqrt{3}}{2}\sigma_{0}+i\frac{1}{2}\sigma_{z}\Big)$ & $\tau_{x}$ & 
    $\begin{pmatrix}
      1/2 &  -\sqrt{3}/2 & 0 \\
       \sqrt{3}/2 & 1/2 & 0 \\
       0 &  0 & 1 \\
     \end{pmatrix}$\\
 $x$ reflection & $\pm\sigma_{x}$ & $\tau_{x}$ & 
    $\begin{pmatrix}
      -1 & 0 & 0 \\
       0 & 1 & 0 \\
       0 & 0 & 1 \\
     \end{pmatrix}$\\
 $y$ reflection & $\pm\sigma_{y}$ & $\tau_{0}$ & 
    $\begin{pmatrix}
       1 & 0 & 0 \\
       0 & -1 & 0 \\
       0 & 0 & 1 \\
     \end{pmatrix}$\\
 \hline
\end{tabular}
\end{table}
%%%%%%%%%%%%%%%%%%%%%%%%%%%%%%%%%%%%%%%%%%%%%%%%%%%%%%
%
%%%%%%%%%%%%%%%%%%%%%%%%%%%%%%%%%%%%%%%%%%%%%%%%%%%%%%
%
%       T A B.  2
%
%%%%%%%%%%%%%%%%%%%%%%%%%%%%%%%%%%%%%%%%%%%%%%%%%%%%%%
\begin{table}[h!]
 \caption{\label{tab:symm_f}List of constraints on $\gamma^{\nu}$ due to the symmetries listed in Tab.\,\ref{tab:symm_op}. The $\gamma^{\nu}$ are hermitian $N$x$N$ matrices, where $N$ is the number of orbitals per unit cell. Thus, they are real if $N=1$. For $N>1$ it is important to note that the complex conjugation of the time reversal operator also acts on $\gamma$. For the orbital degree of freedom in each symmetry operation, \ie the subscript $\mu$ in $\fm{\mu}{\nu}$, we have either exchange of orbitals, $\tau_{x}$, or the identity, $\tau_{0}$. Exchange means all $\fm{yz}{\nu}$ gain an additional factor $-1$.}
 \renewcommand{\arraystretch}{1.2}
 \centering\begin{tabular}{c|l|l}
 \hline
 Symmetry & Conditions for $\gamma^{\nu}$ & Conditions for $\fm{\mu}{\nu}$    \\
 \hline\hline
 \multirow{2}{*}{Time reversal}  &
    $\gamma^{0*}(-k)=\gamma^{0}(k),$ & Even:\ $\fm{0xz}{0},\ \fm{y}{xyz}$\\
  & $(\gamma^{x,y,z})^{*}(-k)=-\gamma^{x,y,z}(k)$ & Odd:\ $\fm{y}{0},\ \fm{0xz}{xyz}$    \\[2pt]
\hline
 \multirow{2}{*}{Inversion}      &
    \multirow{2}{*}{$\tau_{x}\gamma^{\nu}(-k)\tau_{x}=\gamma^{\nu}(k)$} & Even:\ $\fm{0}{\nu},\ \fm{x}{\nu}$  \\
    & & Odd:\ $\fm{y}{\nu},\ \fm{z}{\nu}$ \\[2pt]
\hline\hline
 \multirow{2}{*}{$C_{2}$ rotation} & $\tau_{x}\gamma^{0z}(-k)\tau_{x}=\gamma^{0z}(k)$ & Even:\ $\fm{0x}{0z},\ \fm{yz}{xy}$ \\
    & $\tau_{x}\gamma^{xy}(-k)\tau_{x}=-\gamma^{xy}(k)$ & Odd:\ $\fm{0x}{xy},\ \fm{yz}{0z}$\\[2pt]
\hline
 \multirow{2}{*}{$C_{3}$ rotation} & $\gamma^{0z}(C_{3}k)=\gamma^{0z}(k)$ & Same for $\fm{\mu}{0z}$ \\
    & $\gamma^{x}(C_{3}k)=-\frac{1}{2}\gamma^{x}(k),\ \gamma^{y}(C_{3}k)=-\frac{\sqrt{3}}{2}\gamma^{y}(k)$ & Same for $\fm{\mu}{x},\ \fm{\mu}{y}$ \\[2pt]
\hline
 \multirow{2}{*}{$C_{6}$ rotation} & $\gamma^{0z}(C_{6}k)=\gamma^{0z}(k)$ & Even: $\fm{0x}{0z}$, Odd: $\fm{yz}{0z}$ (under $C_{6}$) \\
    & $\gamma^{x}(C_{6}k)=\frac{1}{2}\gamma^{x}(k),\ \gamma^{y}(C_{6}k)=-\frac{\sqrt{3}}{2}\gamma^{y}(k)$ & Same for $\fm{0x}{x,y}$, factor $-1$ for $\fm{yz}{x,y}$  \\[2pt]
\hline
 \multirow{2}{*}{$x$ reflection}   & $\tau_{x}\gamma^{0x}(-k_{x},k_{y})\tau_{x}=\gamma^{0x}(k_{x},k_{y})$ & Even in $k_{x}$:\ $\fm{0x}{0x},\ \fm{yz}{yz}$ \\
    & $\tau_{x}\gamma^{yz}(-k_{x},k_{y})\tau_{x}=-\gamma^{yz}(k_{x},k_{y})$ & Odd in $k_{x}$:\ $\fm{0x}{yz},\ \fm{yz}{x0}$ \\[2pt]
\hline
 \multirow{2}{*}{$y$ reflection}   & $\gamma^{0y}(k_{x},-k_{y})=\gamma^{0y}(k_{x},k_{y})$ & Even in $k_{y}$:\ $\fm{\mu}{0y}$ \\
    & $\gamma^{xz}(k_{x},-k_{y})=-\gamma^{xz}(k_{x},k_{y})$ & Odd in $k_{y}$:\ $\fm{\mu}{xz}$ \\[2pt]
\hline
\end{tabular}
\end{table}
%%%%%%%%%%%%%%%%%%%%%%%%%%%%%%%%%%%%%%%%%%%%%%%%%%%%%%
%
It is more convenient to check the symmetries of $\fm{\mu}{\nu}$ than of $\gamma^{\nu}$ itself once we have more than one orbital per unit cell, since the Pauli matrices $\tau^{\nu}$ complicate things. We have
\begin{equation}
 \begin{pmatrix}
  \gamma^{0} \\
  \gamma^{x} \\
  \gamma^{y} \\
  \gamma^{z} \\
 \end{pmatrix}=
 \begin{pmatrix}
  \fm{0}{0} & \fm{x}{0} & \fm{y}{0} & \fm{z}{0} \\
  \fm{0}{x} & \fm{x}{x} & \fm{y}{x} & \fm{z}{x} \\
  \fm{0}{y} & \fm{x}{y} & \fm{y}{y} & \fm{z}{y} \\
  \fm{0}{z} & \fm{x}{z} & \fm{y}{z} & \fm{z}{z} \\
 \end{pmatrix}
 \begin{pmatrix}
  \tau^{0} \\
  \tau^{x} \\
  \tau^{y} \\
  \tau^{z} \\
 \end{pmatrix},
\end{equation}
where we give conditions for the $\fm{\mu}{\nu}$ to conserve certain symmetries in Tab.\,\ref{tab:symm_f}. Thus, we can  obtain the factor $D_{\mu\nu}$ in
\begin{equation}
 \fm{\mu}{\nu}(\hat{O}_{r}k)=D_{\mu\nu}\fm{\mu}{\nu}(k)
\end{equation}
arising from the given symmetry under the action of $\hat{O}_{r}$ on $k$. Note that $D_{\mu\nu}$ is mostly just $\pm1$, except for rotations which are not a multiple of $\pi/2$ and reflections that are not in $x$ or $y$ direction. In the following is a list of the factors $D_{\mu\nu}$ for several symmetries:
\begin{align}
 &{\text{Time reversal:}\ }\begin{pmatrix}
             + & + & - & + \\
             - & - & + & - \\
             - & - & + & - \\
             - & - & + & - \\
            \end{pmatrix} \ ,\hspace{4mm}
 {\text{Inversion:}\ }\begin{pmatrix}
             + & + & - & - \\
             + & + & - & - \\
             + & + & - & - \\
             + & + & - & - \\
            \end{pmatrix} \ ,\nonumber\\
\label{eq:symm_f}
 &x\text{ reflection:}\ \begin{pmatrix}
             + & + & - & - \\
             + & + & - & - \\
             - & - & + & + \\
             - & - & + & + \\
            \end{pmatrix} \ ,\hspace{4mm}
 y\text{ reflection:}\ \begin{pmatrix}
             + & + & + & + \\
             - & - & - & - \\
             + & + & + & + \\
             - & - & - & - \\
            \end{pmatrix} \ ,\\
 &C_{2}:\ \begin{pmatrix}
             + & + & - & - \\
             - & - & + & + \\
             - & - & + & + \\
             + & + & - & - \\
            \end{pmatrix}\ , \hspace{4mm}
 C_{3}:\ \begin{pmatrix}
             + & + & + & + \\
             -\frac{1}{2} & -\frac{1}{2} & -\frac{1}{2} & -\frac{1}{2} \\
             -\frac{\sqrt{3}}{2} & -\frac{\sqrt{3}}{2} & -\frac{\sqrt{3}}{2} & -\frac{\sqrt{3}}{2} \\
             + & + & + & + \\
            \end{pmatrix}\ , \hspace{8mm}
 C_{6}:\ \begin{pmatrix}
             + & + & - & - \\
             +\frac{1}{2} & +\frac{1}{2} & -\frac{1}{2} & -\frac{1}{2} \\
             -\frac{\sqrt{3}}{2} & -\frac{\sqrt{3}}{2} & +\frac{\sqrt{3}}{2} & +\frac{\sqrt{3}}{2} \\
             + & + & - & - \\
            \end{pmatrix} \ . \nonumber
\end{align}
\section{Hamiltonian and Bloch matrix}
\label{subsec:Hamilton_Bloch}
\begin{figure}[t]
 \centering\includegraphics[width=0.25\textwidth]{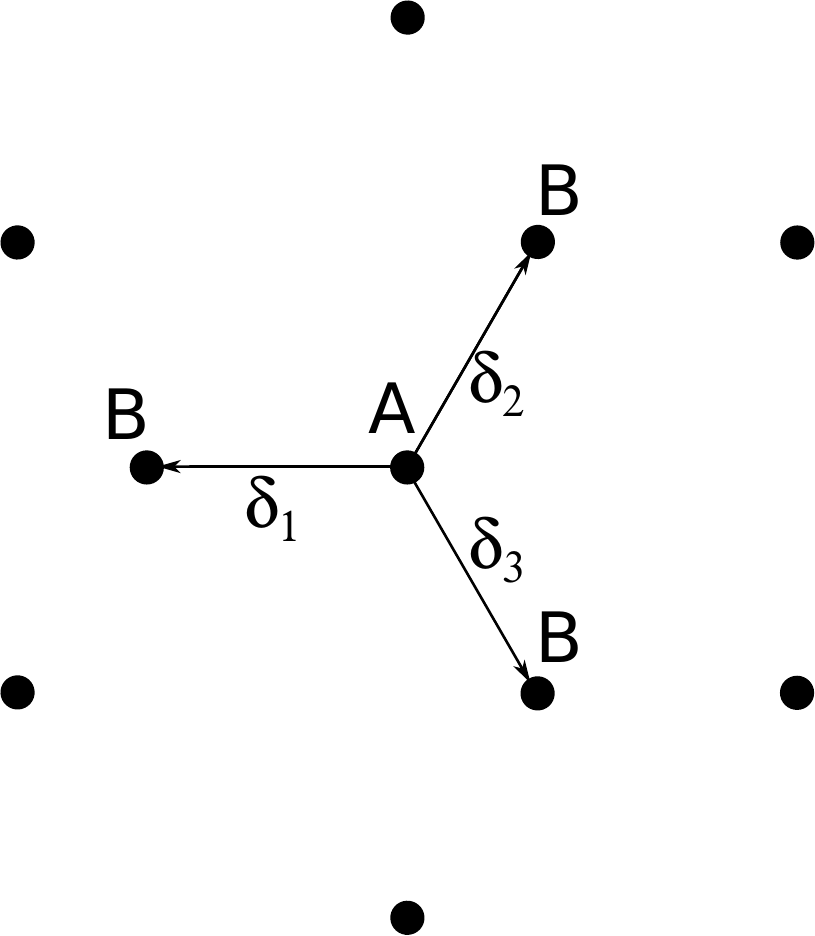}
 \caption{\label{fig:lattice}Hexagonal lattice structure. Neighboring lattice sites are from different sublattices, A and B. The connecting vectors to the three nearest neighbors, $\delta_{1}$, $\delta_{2}$, and $\delta_{3}$, are shown.}
\end{figure}

The non-interacting tight-binding Hamiltonian is in general given by
\begin{equation}
 H_{0}=t_{1}\sum_{\langle i,j\rangle}\sum_{s}c_{is}^{\dagger}c_{js}^{\pd}+t_{2}\sum_{\langle i,j\rangle_{2}}\sum_{s}c_{is}^{\dagger}c_{js}^{\pd}.
\end{equation}
The fundamental lattice structure of all models which will be considered here is that of graphene, which is shown in Fig.\,\ref{fig:lattice}. We have two sites, A and B, per unit cell. The lattice consists of two triangular sublattices A and B, where the nearest neighbor to each lattice site is from the other sublattice. In the following, we will establish the Bloch matrix for each model. The crystal lattice and all hopping and onsite terms for each model are shown in Fig.\,1 of the main text.%\ref{fig:systems}.
%
\iffalse
%%%%%%%%%%%%%%%%%%%%%%%%%%%%%%%%%%%%%%%%%%%%%%%%%%%%%%%%%%%%%%
%           F I G. 1
%%%%%%%%%%%%%%%%%%%%%%%%%%%%%%%%%%%%%%%%%%%%%%%%%%%%%%%%%%%%%%
\begin{figure}[t]
 \includegraphics[width=0.8\columnwidth]{hex_models_lattice_v11.pdf}
 \caption{\label{fig:systems}
 Hoppings and onsite potentials of the four honeycomb lattice systems studied in this paper. Atoms on sublattice $A$ ($B$) are indicated by full (empty) circles.
 (a) Graphene model with nearest and next-nearest neighbor hoppings. 
 (b) Semenoff model with staggered onsite potential $M$. 
 (c) Antiferromagnetic Zeeman model, which is similar to Semenoff model with the difference that the onsite potential $Z$ changes sign for different spins.
 (d) Kane-Mele model, which is similar to the Haldane model, but with an additional sign change for different spins.} 
\end{figure}
%%%%%%%%%%%%%%%%%%%%%%%%%%%%%%%%%%%%%%%%%%%%%%%%%%%%%%%%%%%%%%
\fi
%
%
%
\subsubsection{Graphene}
For graphene (Gr), we consider nearest-neighbor hopping $t_{1}$ and second-neighbor hopping $t_{2}$. The Bloch matrix of this system is given by
\begin{align}
 \hat{h}_{\text{Gr}}&=\begin{pmatrix}
             T_{3}  &   T_{1}-iT_{2}   &   0   &   0 \\
             T_{1}+iT_{2}  &   T_{3}   &   0   &   0 \\
             0  &   0   &   T_{3}   &   T_{1}-iT_{2} \\
             0  &   0   &   T_{1}+iT_{2}   &   T_{3} \\
            \end{pmatrix} \\
            &=\sigma_{0}\otimes(T_{3}\tau_{0}+T_{1}\tau_{x}+T_{2}\tau_{y}),\\
            T_{1}&={\rm Re}\big[t_{1}\sum_{i}e^{k\cdot\delta_{i}}\big]=t_{1}\bigg[\cos(k_{x})+2\cos\bigg(\frac{1}{2}k_{x}\bigg)\cos\bigg(\frac{\sqrt{3}}{2}k_{y}\bigg)\bigg],\\
            T_{2}&=-{\rm Im}\big[t_{1}\sum_{i}e^{k\cdot\delta_{i}}\big]=t_{1}\bigg[\sin(k_{x})-2\sin\bigg(\frac{1}{2}k_{x}\bigg)\cos\bigg(\frac{\sqrt{3}}{2}k_{y}\bigg)\bigg],\\
            T_{3}&=-t_{2}\left[2\cos(\sqrt{3}k_{y})+4\cos\bigg(\frac{\sqrt{3}}{2}k_{y}\bigg)\cos\bigg(\frac{3}{2}k_{x}\bigg)\right],
\end{align}
where we set the length $|\delta_{i}|=1$.
The functions $T_{1}$, $T_{2}$, and $T_{3}$ have the following parities:
\begin{equation}
 T_{1}(k) = T_{1}(-k),\hspace{4mm}T_{2}(k) = -T_{2}(-k),\hspace{4mm}T_{3}(-k)=T_{3}(k).
\end{equation}
The vector $\gamma^{\nu}$ is given by
\begin{align}
\begin{pmatrix}
 \gamma^{0} \\ \gamma^{x} \\ \gamma^{y} \\ \gamma^{z} \\ 
\end{pmatrix}=
 \begin{pmatrix}
  T_{3} & T_{1} & T_{2} & 0 \\
  0 & 0 & 0 & 0 \\
  0 & 0 & 0 & 0 \\
  0 & 0 & 0 & 0 \\
 \end{pmatrix}
\begin{pmatrix}
 \tau^{0} \\ \tau^{x} \\ \tau^{y} \\ \tau^{z} \\ 
\end{pmatrix}.
\end{align}
This satisfies the conditions for all symmetries in Tab.\,\ref{tab:symm_op}.
\subsubsection{Semenoff insulator}
The Semenoff insulator\,\cite{semenoff84prl2449} (Se) introduces a staggered sublattice potential, $M$, \ie we add a constant energy to the Hamiltonian which is positive on sublattice A and negative on sublattice B. This yields for the non-interacting Hamiltonian
\begin{align}
 H_{0}&=t_{1}\sum_{\langle i,j\rangle}\sum_{s}c_{is}^{\dagger}c_{js}^{\pd}+M\sum_{i,s}(-1)^{\varepsilon_{i}}c_{is}^{\dagger}c_{is}^{\pd},\\
 \varepsilon_{i}&=\begin{cases}
                   0 & {\text{on\ sublattice\ A}},\\
                   1 & {\text{on\ sublattice\ B}}.\\
                  \end{cases}
\end{align}
The Bloch matrix takes the form
\begin{align}
 \hat{h}_{\text{Se}}&=\begin{pmatrix}
             M  &   T_{1}-iT_{2}   &   0   &   0 \\
             T_{1}+iT_{2}  &   -M   &   0   &   0 \\
             0  &   0   &   M   &   T_{1}-iT_{2} \\
             0  &   0   &   T_{1}+iT_{2}   &   -M \\
            \end{pmatrix} \\
            &=\sigma_{0}\otimes(T_{1}\tau_{x}+T_{2}\tau_{y}+M\tau_{z}),\\
            T_{1}&={\rm Re}\big[t_{1}\sum_{i}e^{k\cdot\delta_{i}}\big]=t_{1}\bigg[\cos(k_{x})+2\cos\bigg(\frac{1}{2}k_{x}\bigg)\cos\bigg(\frac{\sqrt{3}}{2}k_{y}\bigg)\bigg],\\
            T_{2}&=-{\rm Im}\big[t_{1}\sum_{i}e^{k\cdot\delta_{i}}\big]=t_{1}\bigg[\sin(k_{x})-2\sin\bigg(\frac{1}{2}k_{x}\bigg)\cos\bigg(\frac{\sqrt{3}}{2}k_{y}\bigg)\bigg].
\end{align}
The functions $T_{1}$, $T_{2}$, and $M$ have the following properties:
\begin{align}
 T_{1}(k) &= T_{1}(-k), & T_{2}(k) &= -T_{2}(-k), &  M(k)&=M(-k)={\text{const}}.
\end{align}
The vector $\gamma^{\nu}$ is given by
\begin{align}
\begin{pmatrix}
 \gamma^{0} \\ \gamma^{x} \\ \gamma^{y} \\ \gamma^{z} \\ 
\end{pmatrix}=
 \begin{pmatrix}
  0 & T_{1} & T_{2} & M \\
  0 & 0 & 0 & 0 \\
  0 & 0 & 0 & 0 \\
  0 & 0 & 0 & 0 \\
 \end{pmatrix}
\begin{pmatrix}
 \tau^{0} \\ \tau^{x} \\ \tau^{y} \\ \tau^{z} \\ 
\end{pmatrix}.
\end{align}
The term $\sigma_{0}\otimes M\tau_{z}$ breaks inversion symmetry, since $M(k)=M(-k)$ ($M$ is a constant) and thus $\tau_{x}M\tau_{z}\tau_{x}=-M\tau_{z}$. It also breaks every symmetry with $\hat{O}_{o}=\tau_{x}$, since the staggered potential breaks the symmetry between the sublattices A and B, namely $C_{2}$, $C_{6}$, and $x$ reflection.
\subsubsection{Antiferromagnetic order}
A system with an additional antiferromagnetic Zeeman field (AF) yields the following Hamiltonian:
\begin{align}
 H_{0}&=
 t_{1}\sum_{\langle i,j\rangle}\sum_{s}c_{is}^{\dagger}c_{js}^{\pd}
 +Z\sum_{i}\sum_{ss'}(-1)^{\varepsilon_{i}}c_{is}^{\dagger}s^{z}_{ss'}c_{is'}^{\pd},\\
 \varepsilon_{i}&=\begin{cases}
                   0 & {\text{on\ sublattice\ A}},\\
                   1 & {\text{on\ sublattice\ B}}.\\
                  \end{cases}
\end{align}
The resulting Bloch matrix is given by
\begin{align}
 \hat{h}_{\text{AF}}&=\begin{pmatrix}
             Z  &   T_{1}-iT_{2}   &   0   &   0 \\
             T_{1}+iT_{2}  &   -Z   &   0   &   0 \\
             0  &   0   &   -Z   &   T_{1}-iT_{2} \\
             0  &   0   &   T_{1}+iT_{2}   &   Z \\
            \end{pmatrix} \\
            &=\sigma_{0}\otimes(T_{1}\tau_{x}+T_{2}\tau_{y})+\sigma_{z}\otimes Z\tau_{z},\\
            T_{1}&={\rm Re}\big[t_{1}\sum_{i}e^{k\cdot\delta_{i}}\big]=t_{1}\bigg[\cos(k_{x})+2\cos\bigg(\frac{1}{2}k_{x}\bigg)\cos\bigg(\frac{\sqrt{3}}{2}k_{y}\bigg)\bigg],\\
            T_{2}&=-{\rm Im}\big[t_{1}\sum_{i}e^{k\cdot\delta_{i}}\big]=t_{1}\bigg[\sin(k_{x})-2\sin\bigg(\frac{1}{2}k_{x}\bigg)\cos\bigg(\frac{\sqrt{3}}{2}k_{y}\bigg)\bigg].
\end{align}
The functions $T_{1}$ and $T_{2}$ have the following properties:
\begin{align}
 T_{1}(-k) &= T_{1}(k), & T_{2}(-k) &= -T_{2}(k), & Z(-k) &= Z(k)={\text{const}}.
\end{align}
The vector $\gamma^{\nu}$ is given by
\begin{align}
\begin{pmatrix}
 \gamma^{0} \\ \gamma^{x} \\ \gamma^{y} \\ \gamma^{z} \\ 
\end{pmatrix}=
 \begin{pmatrix}
  0 & T_{1} & T_{2} & 0 \\
  0 & 0 & 0 & 0 \\
  0 & 0 & 0 & 0 \\
  0 & 0 & 0 & Z \\
 \end{pmatrix}
\begin{pmatrix}
 \tau^{0} \\ \tau^{x} \\ \tau^{y} \\ \tau^{z} \\ 
\end{pmatrix}.
\end{align}
The term $\sigma_{z}\otimes Z\tau_{z}$ breaks time reversal symmetry, inversion symmetry, and every symmetry with $\hat{O}_{o}=\tau_{x}$, which are $C_{2}$, $C_{6}$, and $x$ reflection.
\subsubsection{Kane-Mele model}
The Kane-Mele model (KM) presents the quantum spin hall effect in a time reversal invariant system\,\cite{kane-05prl146802,kane-05prl226801}. The Hamiltonian reads
\begin{align}
 H_{0}&=
 t_{1}\sum_{\langle i,j\rangle}\sum_{s}c_{is}^{\dagger}c_{js}^{\pd}
 +it_{2}\sum_{\langle\langle i,j\rangle\rangle}\sum_{ss'}\nu_{ij}c_{is}^{\dagger}\tau^{z}_{ss'}c_{js'}^{\pd}
 ,\\
 \nu_{ij}&=\begin{cases}
           -1 & {\text{clockwise}},\\
           +1 & {\text{counter\ clockwise}}.\\
          \end{cases}
\end{align}
The resulting Bloch matrix is given by
\begin{align}
 \hat{h}_{\text{KM}}&=\begin{pmatrix}
             \Omega  &   T_{1}-iT_{2}   &   0   &   0 \\
             T_{1}+iT_{2}  &   -\Omega   &   0   &   0 \\
             0  &   0   &   -\Omega   &   T_{1}-iT_{2} \\
             0  &   0   &   T_{1}+iT_{2}   &   \Omega \\
            \end{pmatrix} \\
            &=\sigma_{0}\otimes(T_{1}\tau_{x}+T_{2}\tau_{y})+\sigma_{z}\otimes\Omega\tau_{z},\\
            T_{1}&={\rm Re}\big[t_{1}\sum_{i}e^{k\cdot\delta_{i}}\big]=t_{1}\bigg[\cos(k_{x})+2\cos\bigg(\frac{1}{2}k_{x}\bigg)\cos\bigg(\frac{\sqrt{3}}{2}k_{y}\bigg)\bigg],\\
            T_{2}&=-{\rm Im}\big[t_{1}\sum_{i}e^{k\cdot\delta_{i}}\big]=t_{1}\bigg[\sin(k_{x})-2\sin\bigg(\frac{1}{2}k_{x}\bigg)\cos\bigg(\frac{\sqrt{3}}{2}k_{y}\bigg)\bigg],\\
            \Omega&=2t_{2}\bigg[2\cos\bigg(\frac{3}{2}k_{x}\bigg)\sin\bigg(\frac{\sqrt{3}}{2}k_{y}\bigg)-\sin(\sqrt{3}k_{y})\bigg].
\end{align}
The functions $T_{1}$, $T_{2}$, and $\Omega$ have the following parities:
\begin{align}
 T_{1}(-k) &= T_{1}(k), & T_{2}(-k) &= -T_{2}(k), &
 \Omega(-k)&=-\Omega(k).
\end{align}
The vector $\gamma^{\nu}$ is given by
\begin{align}
\begin{pmatrix}
 \gamma^{0} \\ \gamma^{x} \\ \gamma^{y} \\ \gamma^{z} \\ 
\end{pmatrix}=
 \begin{pmatrix}
  0 & T_{1} & T_{2} & 0 \\
  0 & 0 & 0 & 0 \\
  0 & 0 & 0 & 0 \\
  0 & 0 & 0 & \Omega \\
 \end{pmatrix}
\begin{pmatrix}
 \tau^{0} \\ \tau^{x} \\ \tau^{y} \\ \tau^{z} \\ 
\end{pmatrix}\ .
\end{align}
This conserves all symmetries given in Tab.\,\ref{tab:symm_op}.
\subsubsection{Summary}
Summarizing, all symmetries of the four considered models are given in Tab.\,I of the main text.%\ref{tab:summ_symm}.

\section{Bandstructure and eigenvectors}
The full energy spectrum, $\xi(\vec{k})$, of the non-interacting system is obtained by diagonalizing $\hat{h}$ via a unitary transformation,
\begin{align}
\label{eq:total_spectrum}
 \xi(\hat{k})=&\ \hat{U}^{\dagger}(\vec{k})\,\hat{h}(\vec{k})\,\hat{U}(\vec{k})\\
 =&\begin{pmatrix}
         E(\vec{k},-) & 0 & 0 & 0 \\
         0 & E(\vec{k},-) & 0 & 0 \\
         0 & 0 & E(\vec{k},+) & 0 \\
         0 & 0 & 0 & E(\vec{k},+) \\
        \end{pmatrix},\nonumber\\[10pt]
\end{align}
with the band index $n=\pm1$.
Note that all four systems preserve spin degeneracy, and therefore the energy does not depend on the pseudo-spin quantum number $\tilde{s}$. In other words, $E(\vec{k},n,\tilde{s})\equiv E(\vec{k},n)$. The energy spectra for the four models are given by
\begin{equation}
 E(\vec{k},n)=\,\Lambda_{0}(\vec{k})+n\sqrt{T_{1}^{2}(\vec{k})+T_{2}^{2}(\vec{k})+\Lambda_{1}^{2}(\vec{k})}\ , \nonumber
 \end{equation}
where $T_{1}(\vec{k})$ and $T_{2}(\vec{k})$ are given by the nearest-neighbor hopping, and $\Lambda_{0}(\vec{k})$ and $\Lambda_{1}(\vec{k})$ are model dependent and are given by
\begin{align}
 \text{Gr :}\ \ \Lambda_{0}(\vec{k})&=T_{3}(\vec{k})=-t_{2}\left[2\cos(\sqrt{3}k_{y})+4\cos\bigg(\frac{\sqrt{3}}{2}k_{y}\bigg)\cos\bigg(\frac{3}{2}k_{x}\bigg)\right]\ ,\hspace{4mm}\Lambda_{1}(\vec{k})=0\ ,\\
 \text{Se :}\ \ \Lambda_{0}(\vec{k})&=0\ ,\hspace{4mm}\Lambda_{1}(\vec{k})=M=\text{const}\ ,\\[2mm]
 \text{AF :}\ \ \Lambda_{0}(\vec{k})&=0\ ,\hspace{4mm}\Lambda_{1}(\vec{k})=Z=\text{const}\ ,\\
 \text{KM :}\ \ \Lambda_{0}(\vec{k})&=0\ ,\hspace{4mm}\Lambda_{1}(\vec{k})=\Omega(\vec{k})=2t_{2}\bigg[2\cos\bigg(\frac{3}{2}k_{x}\bigg)\sin\bigg(\frac{\sqrt{3}}{2}k_{y}\bigg)-\sin(\sqrt{3}k_{y})\bigg]\ .
\end{align}

The corresponding eigenvectors of $\hat{h}(\vec{k})$, which make up the columns in $\hat{U}(\vec{k})$, are given in Tab.\,\ref{tab:eigenvectors}, where
\begin{align}
 e^{i\phi}=\frac{T_{1}(\vec{k})+iT_{2}(\vec{k})}{\sqrt{T_{1}^{2}(\vec{k})+T_{2}^{2}(\vec{k})}}\ ,\hspace{4mm}\theta_{n}^{x}=\sqrt{1+x/E(\vec{k},n)}\ .
\end{align}
\begin{table}[b]
\caption{\label{tab:eigenvectors}Eigenvectors of the Bloch matrix for the four models.}
\renewcommand{\arraystretch}{1.5}
\centering\begin{tabular}{c|c|c|c|c}
 \hline
 Model & Gr & Se & AF & KM \\
 \hline\hline
 $v^{n\tilde{\dw}}$ &
    $\frac{1}{\sqrt{2}}
    \begin{pmatrix}
        0 \\
        0 \\
        1 \\
        n e^{i\phi} \\
    \end{pmatrix}$ &
    $\frac{1}{\sqrt{2}}
    \begin{pmatrix}
        0 \\
        0 \\
        \theta_{n}^{M} \\
        n e^{i\phi}\theta_{-n}^{M} \\
    \end{pmatrix}$ &
    $\frac{1}{\sqrt{2}}
    \begin{pmatrix}
        0 \\
        0 \\
        \theta_{-n}^{Z} \\
        n e^{i\phi}\theta_{n}^{Z} \\
    \end{pmatrix}$ &
    $\frac{1}{\sqrt{2}}
    \begin{pmatrix}
        0 \\
        0 \\
        \theta_{-n}^{\Omega} \\
        n e^{i\phi}\theta_{n}^{\Omega} \\
    \end{pmatrix}$ \\
    \hline\hline
 $v^{n\tilde{\up}}$ &
    $\frac{1}{\sqrt{2}}
    \begin{pmatrix}
        1 \\
        n e^{i\phi} \\
        0 \\
        0 \\
    \end{pmatrix}$ &
    $\frac{1}{\sqrt{2}}
    \begin{pmatrix}
        \theta_{n}^{M} \\
        n e^{i\phi}\theta_{-n}^{M} \\
        0 \\
        0 \\
    \end{pmatrix}$ &
    $\frac{1}{\sqrt{2}}
    \begin{pmatrix}
        \theta_{n}^{Z} \\
        n e^{i\phi}\theta_{-n}^{Z} \\
        0 \\
        0 \\
    \end{pmatrix}$ &
    $\frac{1}{\sqrt{2}}
    \begin{pmatrix}
        \theta_{n}^{\Omega} \\
        n e^{i\phi}\theta_{-n}^{\Omega} \\
        0 \\
        0 \\
    \end{pmatrix}$ \\
    \hline
\end{tabular}
\end{table}
\section{Weak Coupling RG}
\label{subsec:WCRG}
To find the leading superconducting instabilities for the models outlined above, we use the WCRG method. This method has been discussed previously for spin-invariant systems and systems with lifted spin degeneracy\,\cite{raghu_superconductivity_2010,raghu_effects_2012,wolf_unconventional_2018,wolf_spin-orbit_2020}.
Instead of a full derivation, we will just give a brief summary while pointing out the relevant differences.

In the following, we use the short notation for momentum, band index, and pseudo spin,
\begin{equation}
 1\equiv \vec{k}_{1},n_{1},\tilde{s}_{1}\ ,\hspace{3mm}\bar{1}\equiv-\vec{k}_{1},n_{1},\tilde{s}_{1}\ ,\hspace{3mm}1'\equiv \vec{k}_{1},n_{1},\tilde{s}_{1}'\ .
\end{equation}

The important quantity for calculating the superconducting instabilities in the WCRG method is the two-particle vertex $\Gamma$ in the Cooper channel.
Since the method explicitly demands weak coupling, we expand $\Gamma$ in orders of the local electron-electron interaction, $U_{0}$, up to second order,
\begin{align}
 \Gamma(2\bar{2}'\bar{1}'1)=&\,\sum_{n=1}^{\infty}U_{0}^{n}\Gamma^{(n)}(2\bar{2}'\bar{1}'1)=U_{0}\Gamma^{(1)}(2\bar{2}'\bar{1}'1)+U_{0}^{2}\Gamma^{(2)}(2\bar{2}'\bar{1}'1)+\dots
\end{align}
The first order, $\Gamma^{(1)}$, is given by\,\cite{vafek_spin-orbit_2011}
\begin{align}
\label{eq:Gamma1}
 \Gamma^{(1)}(2\bar{2}'\bar{1}'1)=&\,M(2\bar{2}'\bar{1}'1)\ ,
\end{align}
where the function $M$ reads
\begin{align}
\begin{aligned}
 M(4321)=&\,
  \sum_{l,s\neq s'}v_{l,s}^{*}(4)v_{l,s}^{\ps}(1)v_{l,s'}^{*}(3)v_{l,s'}^{\ps}(2) -\sum_{l,s\neq s'}v_{l,s}^{*}(4)v_{l,s}^{\ps}(2)v_{l,s'}^{*}(3)v_{l,s'}^{\ps}(1)\ .
\end{aligned}
\end{align}
Note that the first order just suppresses the plain $s$-wave solution due to the repulsive interaction, $U_{0}$.
The second order, $\Gamma^{(2)}$, splits into three topologically distinct parts\,\cite{shankar_renormalization-group_1994,vafek_spin-orbit_2011},
\begin{equation}
\label{eq:Gamma2_tot}
 \Gamma^{(2)}=\,\frac{1}{2}\Gamma_{\text{BCS}}+\Gamma_{\text{ZS}}+\Gamma_{\text{ZS'}}\ .
 \end{equation}
The first contribution is referred to as the BCS diagram and the other two contributions to zero sound (ZS) diagrams\,\cite{shankar_renormalization-group_1994}. The BCS diagram diverges logarithmically, which indicates a phase transition, \ie that the normal state becomes instable towards the superconducting state (superconductivity is the only instability in the weak coupling regime,\cite{kohn_new_1965}). The divergency is treated by renormalization, upon which we obtain the superconducting order parameter from the eigenvalue problem
\begin{align}
  \sum_{\hat{k}_{1}n_{1}\tilde{s}_{1}\tilde{s}_{1}'}g(2\bar{2}'\bar{1}'1)\psi_{i}(\hat{k}_{1}n_{1}\tilde{s}_{1}\tilde{s}_{1}')
  =\lambda_{i}\psi_{i}(\hat{k}_{2}n_{2}\tilde{s}_{2}\tilde{s}_{2}')\ ,
\end{align}
where $\hat{k}$ denote momenta on the Fermi surface, $\lambda_{i}$ are the eigenvalues, and $\psi_{i}$ the corresponding form factors projected onto the Fermi surface. The scaled vertex $g$ is used instead of $\Gamma$ since its eigenvalues renormalize independently, \ie the most negative eigenvalue $\lambda_{\text{min}}$ of $g$ is the leading superconducting instability. A relative measure for the critical temperature is given by
\begin{equation}
 T_{c}\propto e^{-1/|\lambda_{\text{min}}|}=e^{-1/\rho |U_{\text{eff}}|}\ .
\end{equation}

The scaled vertex $g$ is obtained from $\Gamma$ by
\begin{align}
 g(2\bar{2}'\bar{1}'1)&=\tau(2)\big[\Gamma_{\text{ZS}}(2\bar{2}'\bar{1}'1)+\Gamma_{\text{ZS}'}(2\bar{2}'\bar{1}'1)\big]\tau(1)\nonumber\\
 &=\tau(2)\big[\Gamma_{\text{ZS}}(2\bar{2}'\bar{1}'1)-\Gamma_{\text{ZS}}(\bar{2}'2\bar{1}'1)\big]\tau(1)\ ,
\end{align}
where the scaling function $\tau(1)$ is defined for a discretized Fermi surface as
\begin{align}
 \tau(\hat{k},n)=\sqrt{\rho_{n}\frac{l(\hat{k},n)\bar{v}(n)}{S_{F}(n)v(\hat{k},n)}}\ ,\hspace{4mm}
 \frac{1}{\bar{v}(n)}=\sum_{\hat{k}}\frac{l(\hat{k},n)}{S_{F}(n)}\frac{1}{v(\hat{k},n)}\ .
\end{align}
Here, $\rho_{n}$ denotes the density of states at the Fermi level of band $n$, $l(\hat{k},n)$ the distance to the neighboring Fermi surface discretization points, $S_{F}(n)$ the total length of the Fermi surface of band $n$, and $v(\hat{k},n)$ the Fermi velocity of band $n$ at the momentum $\hat{k}$.

With the vertex $\Gamma_{\text{ZS}}$ given by
\begin{align}
 \Gamma_{\rm ZS}(2\bar{2}'\bar{1}'1)=\sum_{\vec{k}_{3}}\sum_{n_{3}n_{4}}&\sum_{\tilde{s}_{3}\tilde{s}_{4}}M(\bar{2}'431)M(234\bar{1}')\frac{f(E(3))-f(E(4))}{E(3)-E(4)}=-\Gamma_{\rm ZS'}(\bar{2}'2\bar{1}'1)\ ,
\end{align}
where $f(E)$ is the Fermi distribution, we obtain for the scaled vertex
\begin{align}
 &g(\hat{k}_{2}n_{2}\tilde{s}_{2}\tilde{s}_{2}',\hat{k}_{1}n_{1}\tilde{s}_{1}\tilde{s}_{1}')=\tau(\hat{k}_{2},n_{2})\Big[
  \Pi( \hat{k}_{2}n_{2}\tilde{s}_{2} \tilde{s}_{2}',\hat{k}_{1}n_{1}\tilde{s}_{1}\tilde{s}_{1}')
 -\Pi(-\hat{k}_{2}n_{2}\tilde{s}_{2}'\tilde{s}_{2} ,\hat{k}_{1}n_{1}\tilde{s}_{1}\tilde{s}_{1}')\Big]\ ,\\[3mm]
\label{eq:Pi}
 &\Pi( \hat{k}_{2}n_{2}\tilde{s}_{2} \tilde{s}_{2}',\hat{k}_{1}n_{1}\tilde{s}_{1}\tilde{s}_{1}')=
 \sum_{\nu,\nu'}\sum_{\tilde{\zeta},\tilde{\zeta}'}\int_{\text{BZ}}\frac{\text{d}^{2}q}{(2\pi)^{2}}\frac{f(E(\vec{q},\nu))-f(E(\vec{Q},\nu'))}{E(\vec{q},\nu)-E(\vec{Q},\nu')}\\
 &\hspace{3cm}\times
 M(-\hat{k}_{2}n_{2}\tilde{s}_{2}',\vec{Q}\nu'\tilde{\zeta}',\vec{q}\nu\tilde{\zeta},\hat{k}_{1}n_{1}\tilde{s}_{1})
 M(\hat{k}_{2}n_{2}\tilde{s}_{2},\vec{q}\nu\tilde{\zeta},\vec{Q}\nu'\tilde{\zeta}',-\hat{k}_{1}n_{1}\tilde{s}_{1}')\ ,\nonumber
\end{align}
where $\vec{Q}=\vec{q}+\hat{k}_{1}+\hat{k}_{2}$ and the integral runs over the whole Brillouin-zone (BZ). Since we have four different spin indices in $g$, there are 16 possible spin channels. However, due to hermiticity and fermionic anti-commutation relations, we are left with six independent spin channels, given by
\begin{align}
 A=&\,g(\hat{k}_{2}n_{2}\tilde{\dw}\tilde{\dw},\hat{k}_{1}n_{1}\tilde{\dw}\tilde{\dw})\ ,\\
 B=&\,g(\hat{k}_{2}n_{2}\tilde{\dw}\tilde{\dw},\hat{k}_{1}n_{1}\tilde{\dw}\tilde{\up})
   =-g(-\hat{k}_{2}n_{2}\tilde{\dw}\tilde{\dw},\hat{k}_{1}n_{1}\tilde{\up}\tilde{\dw})
   =g^{*}(\hat{k}_{1}n_{1}\tilde{\dw}\tilde{\up},\hat{k}_{2}n_{2}\tilde{\dw}\tilde{\dw})
   =-g^{*}(-\hat{k}_{1}n_{1}\tilde{\up}\tilde{\dw},\hat{k}_{2}n_{2}\tilde{\dw}\tilde{\dw})\ ,\\
 C=&\,g(\hat{k}_{2}n_{2}\tilde{\dw}\tilde{\dw},\hat{k}_{1}n_{1}\tilde{\up}\tilde{\up})
   =g^{*}(\hat{k}_{1}n_{1}\tilde{\up}\tilde{\up},\hat{k}_{2}n_{2}\tilde{\dw}\tilde{\dw})\ ,\\
 D=&\,g(\hat{k}_{2}n_{2}\tilde{\dw}\tilde{\up},\hat{k}_{1}n_{1}\tilde{\dw}\tilde{\up})
   =-g(-\hat{k}_{2}n_{2}\tilde{\dw}\tilde{\up},\hat{k}_{1}n_{1}\tilde{\up}\tilde{\dw})
   =-g(\hat{k}_{2}n_{2}\tilde{\up}\tilde{\dw},-\hat{k}_{1}n_{1}\tilde{\dw}\tilde{\up})
   =g(-\hat{k}_{2}n_{2}\tilde{\up}\tilde{\dw},-\hat{k}_{1}n_{1}\tilde{\up}\tilde{\dw})\ ,\\
 E=&\,g(\hat{k}_{2}n_{2}\tilde{\dw}\tilde{\up},\hat{k}_{1}n_{1}\tilde{\up}\tilde{\up})
   =-g(\hat{k}_{2}n_{2}\tilde{\up}\tilde{\dw},-\hat{k}_{1}n_{1}\tilde{\up}\tilde{\up})
   =g^{*}(\hat{k}_{1}n_{1}\tilde{\up}\tilde{\up},\hat{k}_{2}n_{2}\tilde{\dw}\tilde{\up})
   =g^{*}(\hat{k}_{1}n_{1}\tilde{\up}\tilde{\up},-\hat{k}_{2}n_{2}\tilde{\up}\tilde{\dw})\ ,\\
 F=&\,g(\hat{k}_{2}n_{2}\tilde{\up}\tilde{\up},\hat{k}_{1}n_{1}\tilde{\up}\tilde{\up})\ .
\end{align}
In all models studied here, since there are no spin-flips involved in the interaction, the spin channels $B$, $C$, and $E$ vanish. This separates the channels with total pseudo-spin $\tilde{S}=\pm1$ ($A$ and $F$) from the channels with total pseudo-spin $\tilde{S}=0$ ($D$).

The contribution of the eigenvectors to the vertex, as given by the functions $M(\dots)M(\dots)$ in Eq.\,\eqref{eq:Pi}, can be calculated analytically. The result is given by
\begin{align}
%%%%%%%%%%
%
%   Gr
%
%%%%%%%%%%
 \text{Gr:}\hspace{4mm}&\frac{1}{8}\Bigg[1+n_{1}n_{2} \nu \nu'\frac{
    T_{r}^{4}
  }{\sqrt{T_{1}^{2}(k_{1})+T_{2}^{2}(k_{1})}
    \sqrt{T_{1}^{2}(k_{2})+T_{2}^{2}(k_{2})}
    \sqrt{T_{1}^{2}(q)+T_{2}^{2}(q)}
    \sqrt{T_{1}^{2}(Q)+T_{2}^{2}(Q)}}\Bigg]\ ,\\
%%%%%%%%%%
%
%   Se
%
%%%%%%%%%%
 \text{Se:}\hspace{4mm}&\frac{1}{8}\Bigg[1+\frac{
    T_{r}^{4}+M^{4}
    }{E(k_{1},n_{1})E(k_{2},n_{2})E(q,\nu)E(Q,\nu')}
     +\frac{M^{2}}{E(Q,\nu')     E(q,\nu)}
     +\frac{M^{2}}{E(k_{1},n_{1})E(Q,\nu')}\\
  &\hspace{10mm}
     +\frac{M^{2}}{E(k_{1},n_{1})E(q,\nu)}
     +\frac{M^{2}}{E(k_{2},n_{2})E(Q,\nu')}
     +\frac{M^{2}}{E(k_{2},n_{2})E(q,\nu)}
     +\frac{M^{2}}{E(k_{1},n_{1})E(k_{2},n_{2})}\Bigg]\ ,\nonumber\\
%%%%%%%%%%
%
%   AF
%
%%%%%%%%%%
 \text{AF, }A\text{ and }F\hspace{4mm}&\frac{1}{8}\Bigg[1+\frac{T_{r}^{4}+Z^{4}
  }{E(k_{1},n_{1})E(k_{2},n_{2})E(q,\nu)E(Q,\nu')}
     +\frac{Z^{2}}{E(Q,\nu')     E(q,\nu)}
     -\frac{Z^{2}}{E(k_{1},n_{1})E(Q,\nu')}\\
    &\hspace{10mm}
     -\frac{Z^{2}}{E(k_{1},n_{1})E(q,\nu)}
     -\frac{Z^{2}}{E(k_{2},n_{2})E(Q,\nu')}
     -\frac{Z^{2}}{E(k_{2},n_{2})E(q,\nu)}
     +\frac{Z^{2}}{E(k_{1},n_{1})E(k_{2},n_{2})}\Bigg]\ ,\nonumber\\
 \text{AF, }D:\hspace{4mm}&\frac{n_{1}n_{2}}{8}\Bigg[
    \frac{\sqrt{T_{1}^{2}(k_{1})+T_{2}^{2}(k_{1})}\sqrt{T_{1}^{2}(k_{2})+T_{2}^{2}(k_{2})}}{E(k_{1},n_{1})E(k_{2},n_{2})}\Bigg(1-\frac{Z^{2}}{E(q,\nu)E(Q,\nu')}\Bigg)\\
    &\hspace{50mm}+\frac{
 T_{r}^{4}\Big(1-\frac{Z^{2}}{E(k_{1},n_{1})E(k_{2},n_{2})}\Big)
 +i\,T_{i}^{4}\Big(\frac{Z}{E(k_{1},n_{1})}-\frac{Z}{E(k_{2},n_{2})}\Big)}{\sqrt{T_{1}^{2}(k_{1})+T_{2}^{2}(k_{1})}\sqrt{T_{1}^{2}(k_{2})+T_{2}^{2}(k_{2})}E(q,\nu)E(Q,\nu')}\Bigg]\ ,\nonumber
\end{align}
\begin{align}
%%%%%%%%%%
%
%   KM
%
%%%%%%%%%%
 \text{KM, }A\text{ and }F:\hspace{4mm}&\frac{n_{1}n_{2}}{8}\Bigg[
    \frac{\sqrt{T_{1}^{2}(k_{1})+T_{2}^{2}(k_{1})}\sqrt{T_{1}^{2}(k_{2})+T_{2}^{2}(k_{2})}}{E(k_{1},n_{1})E(k_{2},n_{2})}
    \Bigg(1+\frac{\Omega(q)\Omega(Q)}{E(q,\nu)E(Q,\nu')}\Bigg)\\
   &\hspace{48mm}+\frac{T_{r}^{4}\Big(1-\frac{\Omega(k_{1})\Omega(k_{2})}{E(k_{1},n_{1})E(k_{2},n_{2})}\Big)
    +i\,T_{i}^{4}\Big(\frac{\Omega(k_{1})}{E(k_{1},n_{1})}-\frac{\Omega(k_{2})}{E(k_{2},n_{2})}\Big)
 }{\sqrt{T_{1}^{2}(k_{1})+T_{2}^{2}(k_{1})}\sqrt{T_{1}^{2}(k_{2})+T_{2}^{2}(k_{2})}E(q,\nu)E(Q,\nu')}\Bigg]\ ,\nonumber\\
\text{KM, }D:\hspace{4mm}&\frac{1}{8}\Bigg[1
 +\frac{T_{r}^{4}-\Omega(k_{1})\Omega(k_{2})\Omega(q)\Omega(Q)}{E(k_{1},n_{1})E(k_{2},n_{2})E(q,\nu)E(Q,\nu')}
    +\frac{\Omega(Q)\Omega(k_{2})}{E(Q,\nu')E(k_{2},n_{2})}
    +\frac{\Omega(Q)\Omega(k_{1})}{E(Q,\nu')E(k_{1},n_{1})}\\
   &\hspace{12mm}-\frac{\Omega(Q)\Omega(q)}{E(Q,\nu')E(q,\nu)}
    +\frac{\Omega(k_{1})\Omega(k_{2})}{E(k_{1},n_{1})E(k_{2},n_{2})}
    -\frac{\Omega(q)\Omega(k_{2})}{E(q,\nu)E(k_{2},n_{2})}
    -\frac{\Omega(q)\Omega(k_{1})}{E(q,\nu)E(k_{1},n_{1})}
    \Bigg]\ ,\nonumber
\end{align}
where
\begin{align}
     T_{r}^{4}=&
            \big[T_{1}(k_{1})T_{2}(k_{2})+T_{1}(k_{2})T_{2}(k_{1})\big]
            \big[T_{1}(q)T_{2}(Q)-T_{2}(q)T_{1}(Q)\big]\\
  &\hspace{42mm}
           +\big[T_{1}(k_{1})T_{1}(k_{2})-T_{2}(k_{1})T_{2}(k_{2})\big]
            \big[T_{1}(q)T_{1}(Q)+T_{2}(q)T_{2}(Q)\big]\ ,
            \nonumber\\
    T_{i}^{4}=&
            \big[T_{1}(k_{1})T_{2}(k_{2})+T_{1}(k_{2})T_{2}(k_{1})\big]
            \big[T_{1}(q)T_{1}(Q)+T_{2}(q)T_{2}(Q)\big]\\
  &\hspace{42mm}
           +\big[T_{1}(k_{1})T_{1}(k_{2})-T_{2}(k_{1})T_{2}(k_{2})\big]
            \big[T_{2}(q)T_{1}(Q)-T_{1}(q)T_{2}(Q)\big]\ .
            \nonumber
\end{align}
Note that the pseudo-spin indices disappeared. Their only effect is choosing the correct eigenvectors for each spin channel. In the expressions for the eigenvectors themselves, once the pseudo-spin is defined, the pseudo-spin indices do not appear. Furthermore, it is apparent that for AF and KM, the vertices with $\tilde{s}_{z}=0$ are different from the ones with $S_{z}=\pm1$, whereas in all other models there is not difference.

The parameters used for the calculations in the main paper for the small gap case are given in Tab.\,\ref{tab:parameters}
\begin{table}[b]
 \caption{\label{tab:parameters}Parameter sets used for the calculations in the main text for the small gap case. Each line yields almost identical Fermi surfaces between the four models. For the large gap case, the graphene model has $t_{2}=0.35$ and the Kane-Mele model $t_{2}=0.5$ for all values of $\mu$. For the fillings $n$ marked with a star the following changes apply: at $n=1.21$, the KM model has $n=1.20$. At $n=1.25$, the Gr model has $n=1.255$. At $n=1.28$, the Gr model has $n=1.29$.}
  \centering\begin{tabular}{lll|llll}
  \hline
  & & Model   & Gr & Se & AF & KM \\
  Set & $\mu$ & $n$ & \multicolumn{4}{c}{$t_{2}$} \\
  \hline
  1  & 0.6  & 1.06       & $0.017$  & 0.23  & 0.23  & 0.05 \\
  2  & 0.65 & 1.07       & $0.015$  & 0.22  & 0.22  & 0.05 \\
  3  & 0.7  & 1.09       & $0.013$  & 0.21  & 0.21  & 0.05 \\
  4  & 0.75 & 1.11       & $0.011$  & 0.20  & 0.20  & 0.05 \\
  5  & 0.8  & 1.13       & $0.009$  & 0.195 & 0.195 & 0.05 \\
  6  & 0.9  & 1.17       & $0.0055$ & 0.15  & 0.15  & 0.05 \\
  7  & 0.95 & 1.21$^{*}$ & $0.004$  & 0.13  & 0.13  & 0.05 \\
  8  & 0.98 & 1.23       & $0.003$  & 0.11  & 0.11  & 0.05 \\
  9  & 0.99 & 1.24       & $0.0025$ & 0.11  & 0.11  & 0.05 \\
  10 & 1.01 & 1.25$^{*}$ & $0.002$  & 0.10  & 0.10  & 0.05 \\
  11 & 1.02 & 1.26       & $0.001$  & 0.10  & 0.10  & 0.05 \\
  12 & 1.05 & 1.28$^{*}$ & $0.003$  & 0.11  & 0.11  & 0.05 \\
  13 & 1.1  & 1.32       & $0.004$  & 0.12  & 0.12  & 0.05 \\
  14 & 1.2  & 1.37       & $0.004$  & 0.13  & 0.13  & 0.05 \\
  15 & 1.4  & 1.47       & $0.004$  & 0.13  & 0.13  & 0.05 \\
  \hline
 \end{tabular}
\end{table}
\section{Form factors of the superconducting order parameter}
\label{sec:formfactors}
Several examples of appearing formfactors will be given in the following.

\begin{figure}[ht]
 \centering
 \includegraphics[width=0.1\textwidth]{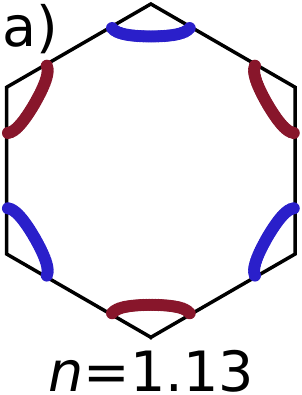}\hspace{2mm}
 \includegraphics[width=0.1\textwidth]{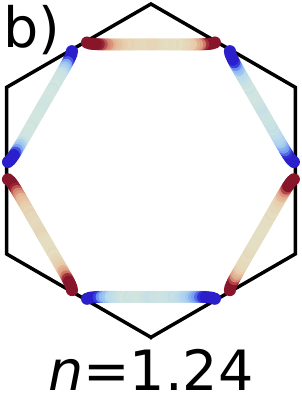}\hspace{2mm}
 \includegraphics[width=0.1\textwidth]{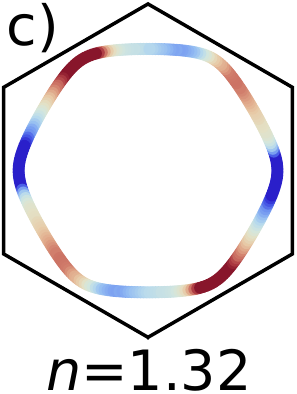}\hspace{2mm}
 \includegraphics[width=0.1\textwidth]{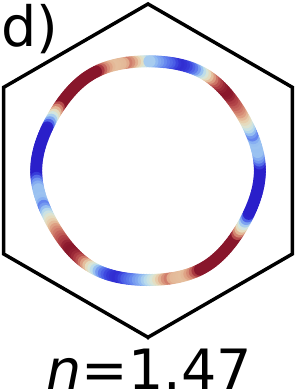}
 \caption{\label{fig:small_gap_formfactors}Examples of formfactors in the small gap regime for several fillings $n$. The chemical potentials $\mu$ are (a) $0.8$, (b) $0.99$, (c) $1.1$, and (d) $1.4$. (a) and (b) show a node-less $f$-wave (irrep $B_{1}$), (c) and (d) a $d+id$-wave (irrep $E_{2}$, only one of the two degenerate $d$-waves is shown).}
\end{figure}

Firstly, in the small gap case, the formfactors of the leading superconducting instability is the same for each of the five models, as discussed in the main text. A few examples of the formfactors projected to the FS are shown in Fig.\,\ref{fig:small_gap_formfactors}. Panel (a) shows a node-less $f$-wave (irrep $B_{1}$) at $\mu=0.8$. Increasing $\mu$ towards the van-Hove singularity, the gap is mostly concentrated around the points on the FS which are close to the $M$ points in the Brillouin zone, where the density of states diverges. This is shown in Fig.\,\ref{fig:small_gap_formfactors}\,(b), where $\mu=0.99$. Panel (c) shows one of the degenerate $d$-wave formfactors of the $E_{2}$ irrep, for $\mu=1.1$. The high number of nodal lines indicates higher harmonics present in this state (cf.~formfactor shown in the main text). At higher chemical potential, shown in panel (d) at $\mu=1.4$, the formfactor becomes even more mixed with higher harmonics.

\begin{figure}[ht]
 \centering\includegraphics[width=0.95\textwidth]{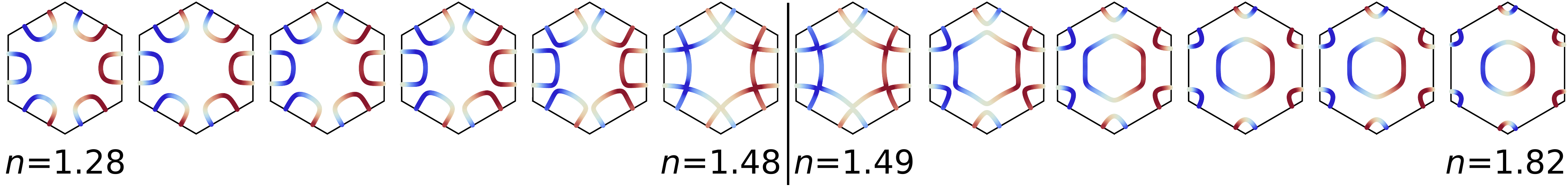}
 \caption{\label{fig:Hd_KM_pwave}Evolution of the $p_x$-wave formfactor in the Kane-Mele model for the large gap case for increasing filling $n$. The vertical line denotes the van-Hove singularity. Note that only the $p_x$-wave is shown, there is always a degenerate $p_y$-wave for each of the shown states.}
\end{figure}

For the Kane-Mele model in the large gap case, \ie for $t_{2}=0.5$, the main text shows a dominating $p+ip$-wave order parameter. Interestingly, this order parameter survives when the chemical potential $\mu$ is increased across the van-Hove singularity. The formfactors projected on the FS for values of $\mu$ between $1.85$ and $2.5$ are shown in Fig.\,\ref{fig:Hd_KM_pwave}, and show a very clean $p$-wave, \ie without contributions of higher harmonics.

\begin{figure}[ht]
 \centering
 \includegraphics[width=0.1\textwidth]{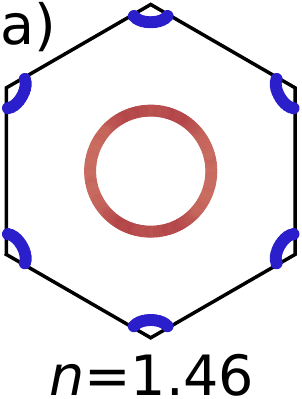}\hspace{2mm}
 \includegraphics[width=0.1\textwidth]{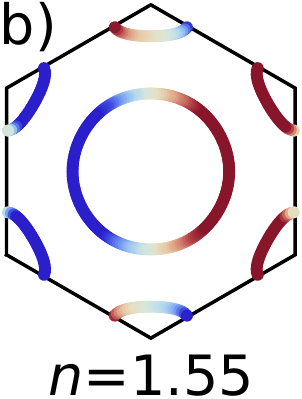}\hspace{2mm}
 \includegraphics[width=0.1\textwidth]{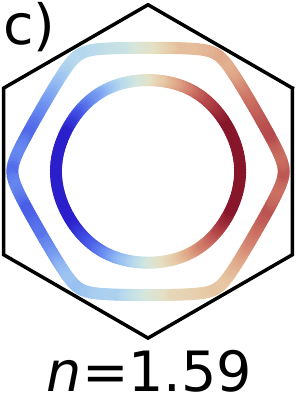}
 \caption{\label{fig:Gr_examples}Examples of the formfactors of the superconducting order parameter of graphene with second neighbor hopping $t_{2}=0.35$ for different fillings $n$. The shown formfactors are from leading superconducting instabilities. (a) shows a node-less extended $s$-wave at $\mu=1.4$, (b) and (c) show one of the two degenerate $p$-wave states at $\mu=1.6$ and $\mu=1.71$, respectively.}
\end{figure}

For the graphene model, we show three formfactors in Fig.\,\ref{fig:Gr_examples}. Here, we show a nodeless extended $s$-wave in Fig.\,\ref{fig:Gr_examples}\,(a), and $p$-waves in (b) and (c).
\section{Comparison of the bandstructures}
\begin{figure*}[h!]
  \centering\includegraphics[width=0.9\columnwidth]{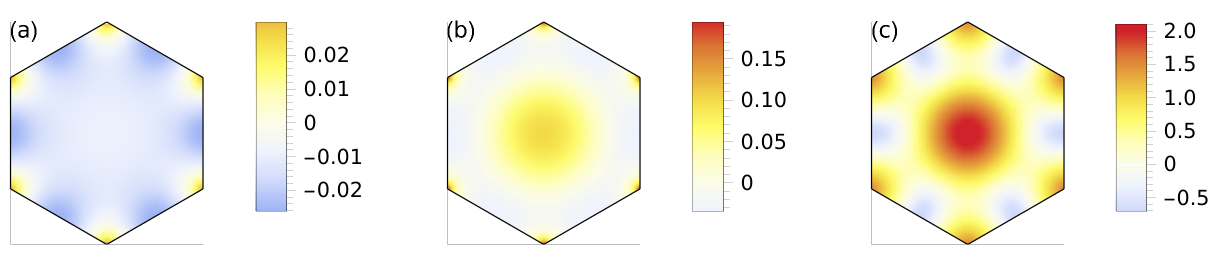}
  \caption{\label{fig:BS_comp}Difference between the bandstructures of the models. We refer to the energy bands of the KM, Se and Gr models as $E_{\rm KM}$, $E_{\rm Se}$ and $E_{\rm Gr}$, respectively. (a) shows $E_{\text{KM}}-E_{\text{Se}}$ in the small gap case for $n=1.06$ and $t_{2,\text{Se}}=0.23$, (b) shows $E_{\text{KM}}-E_{\text{Gr}}$ in the small gap case for $n=1.06$ and $t_{2,\text{Gr}}=0.017$, and (c) shows $E_{\text{KM}}-E_{\text{Gr}}$ in the large gap case for $n_{\text{KM}}=1$ and $n_{\text{Gr}}=0.86$ ($\mu=0$ in both cases).}
\end{figure*}

Here, we extend on our argument that eliminating the effect of the shape of the Fermi surface. In particular, we investigate the differences in the bandstructure between the models. This plays a role in the Lindhard function in Eq.\,\eqref{eq:Pi}, \ie for the fraction
\begin{equation}
 \frac{f(E(\vec{q},\nu))-f(E(\vec{Q},\nu'))}{E(\vec{q},\nu)-E(\vec{Q},\nu')}\ .
\end{equation}
Since we divide by the energy, momenta which are far away from the Fermi surface in terms of energy have negligible impact. In Fig.\,\ref{fig:BS_comp} we show the difference of the bandstructures between the KM, Se, and Gr models. Panel (a) shows the difference between KM and Se for the small gap case, panel (b) between KM and Gr in the small gap case, and panel (c) between KM and Gr in the large gap case. Overall, the KM and Se models are the most similar over the whole BZ, whereas for the comparison with Gr, we see the strongest deviations at the $K$ and $K'$ points. This is to be expected, since in the Gr model the gap closes at these points. In the large gap case, the bandstructures become more different, since the model-specific terms become larger. Around the Fermi surface, however, the difference is still small.

\newpage

\bibliography{wcrg3}